\documentclass[amsmath,amssymb,amsfonts,floatfix
aps,physrev,
twocolumn,groupedaddress,bibnotes]{revtex4-2}
\usepackage[utf8]{inputenc}
\usepackage{graphicx}
\usepackage[]{physpack}

\usepackage{color}
\definecolor{linkcolor}{RGB}{0,75,175}
\usepackage[colorlinks=true,
            linkcolor=linkcolor,
            urlcolor=linkcolor,
            citecolor=linkcolor,
            unicode,
            pdfencoding=auto]{hyperref}

\begin{document}


\title{Crossover from Fast Scrambling to Operator Confinement Tuned by an Auxiliary Qubit}


\author{Ryan Buechele}
\email{buechele.5@osu.edu}
\affiliation{Department of Physics, The Ohio State University, Columbus, Ohio 43210, USA}

\author{Joseph C. Szabo}
\email{jszabo94@gmail.com}
\affiliation{Department of Physics, The Ohio State University, Columbus, Ohio 43210, USA}

\author{Nandini Trivedi}
\email{trivedi.15@osu.edu}
\affiliation{Department of Physics, The Ohio State University, Columbus, Ohio 43210, USA}

\date{\today}

\begin{abstract}
We demonstrate a static, disorder-free spin chain Hamiltonian which, by tuning the coupling to an auxiliary qubit, realizes a crossover between super-ballistic, ancilla-accelerated scrambling and sub-ballistic operator confinement. Our minimal model is the mixed-field Ising chain with a spin-1/2 ancilla coupled to its longitudinal magnetization. The ancilla mediates an effective all-to-all interaction which accelerates operator spreading and entanglement growth when weakly coupled, but rapidly saturates its entanglement and projects spin chain operators into effectively frozen subspaces when the ancilla coupling is strong. We locate this crossover independently through both a divergent peak in the mutual-information saturation time near $\lambda_c N/h \approx 8$ and an exponential suppression of the late-time OTOC growth rate, $\log\alpha \propto 1/\lambda$. Through a Feshbach-Fano projection and Schrieffer-Wolff transformation, we reveal an effective hidden symmetry on the chain which confines operators on the chain for a time exponential in the coupling strength. This reconciles the fast, $\log(N)$ scrambling reported for random-unitary-circuit realizations of the star geometry with the confinement previously found in its time-independent Hamiltonian analog, showing both emerge from a single Hamiltonian family as a function of one dimensionless parameter.
\end{abstract}


\maketitle

\section{\label{sec:Introduction} Introduction}

Operator scrambling and entanglement entropy spreading are unambiguous discriminators of purely quantum mechanical nonequilibrium dynamics, underlying quantum thermalization, dynamical phase transitions, and topological order \cite{Kitaev_topoent2006, wen_topoent2006, Hastings_arealaw2007, Haldane_spectrum2008, pollmann_spectrum2010, Heyl2018, Dag2020}. In the Heisenberg picture, scrambling describes how initially localized operators propagate through noncommuting many-body interactions. In the complementary Schr\"odinger picture, the von Neumann entropy captures how initially classical states develop entangled structure and growing information complexity. Both perspectives have become central to a wave of recent quantum simulation and circuit experiments \cite{kinoshita2006quantum, landsman2019, joshi2020, Mi_scrambling2021, blok2021, mi2022time, Zhu_scramble_2022, weinstein2023scrambling, seki2025simulating, liang_observation_2025}, driving further work at the intersection of quantum chaos, thermalization, and computability, from qubits to black holes \cite{Srednicki_entropy1993, Sekino_2008, Lashkari2013, Maldacena_syk2016, Maldacena_chaos2016, Chen2017, Bohrdt_2017, Rigol2008, lewis-swan2019}.
 
Closed systems exhibit rich scrambling behavior -- frozen \cite{Swingle2017}, fast \cite{Lashkari2013}, or either depending on initial conditions \cite{banuls_strong_2011}, naturally raising the question if and how this behavior survives once the system couples to an external environment \cite{zeng2017prethermal, moessner2017equilibration, Touil2021, Zanardi2021}. Rather than invoking a fully Markovian bath, which assumes weak coupling and a wide separation of time scales between system and bath \cite{mazzola2010phenomenological, cohen1998atom}, a more tractable route traces out an auxiliary subsystem of a larger closed, unitary model to treat the remainder as an open system. This turns the interaction topology and coupling strength of the ``environment'' into a tunable parameter rather than an assumption, directly relevant to systems with an intrinsic auxiliary qubit degree of freedom, such as mechanical or optical modes \cite{chu2017quantum, mirhosseini2020superconducting, kok2007linear, stockill2017phase}.

Extensively scaling, nonlocal couplings generically accelerate scrambling, with a rate that grows with system size \cite{bentsen_treelike2019, Li2020, Belyansky2020}, and random-unitary-circuit (RUC) realizations of a star-graph topology confirm fast, $\log N$-scrambling in that geometry \cite{lucas2019quantum, Harrow2021}; however, Ref.~\cite{lucas2019quantum} shows a time-independent Hamiltonian on the same star geometry instead confines operators on the ancilla with a coherent lifetime. This tension between fast RUC scrambling and slow Hamiltonian-driven confinement on the same star geometry motivates our goal for a time-independent, nonintegrable Hamiltonian which includes both regimes of accelerated scrambling versus confinement as a function of a single parameter.

This report studies a single ancilla qubit coupled to the global magnetization of a spin-1/2 chain in a mixed field,
\begin{equation}
\label{eq:full_ring_star_ham}
\mathsf{H} = J\sum_j S_j^z S_{j+1}^z + \lambda S^z S_A^z - hS^x - h_AS_A^x - gS^z - g_AS_A^z ,
\end{equation}
where $S^\alpha = \sum_j S_j^\alpha =\frac{1}{2}\sum_j \sigma_j^\alpha$. Tracing out the ancilla yields emergent, non-Hermitian dynamics on the chain that we contrast directly against the closed, unitary dynamics of the isolated chain, with the ancilla playing the role of the simplest structured environment. The ancilla coupling strength $\lambda$ acts as a single dial spanning two opposite dynamical regimes within one closed, time-independent, disorder-free Hamiltonian. At weak coupling, the ancilla mediates an effective all-to-all interaction that accelerates scrambling and thermalization. At strong coupling, the ancilla instead rapidly saturates its own entanglement with the chain, aliasing operators that fail to commute with the coupling and imposing an effective hidden symmetry that slows multi-particle entanglement growth and operator complexity, generating disorder-free localization through Hilbert-space fragmentation \cite{sala_ergodicity_2020, yang_hilbertspace_2020, jeyaretnam_hilbert_2025, sriram_fragmented_2026}.

Competition between the noncommuting ancilla coupling and the applied Zeeman field drives this transition. We project out the ancilla in analytic expansions to expose the responsible effective interactions, then illustrate the resulting behavior through numerically exact calculation of entanglement measures and out-of-time-order correlators (OTOCs). The observed confinement echoes two established phenomena for different underlying reasons: driven Floquet systems, where periodic driving imprints an effective symmetry that produces prethermalization and slowed entanglement growth \cite{kyprianidis2021observation, bhakuni2021suppression, sahu2025information}, and the quantum Zeno effect, where repeated projective measurement drives a ballistic-to-sub-ballistic entanglement transition \cite{Li2018, Li2019}. Here, however, a single static, global coupling generates the projection, distinguishing the mechanism from both local disorder/purification networks and periodic driving that conventionally produces such crossovers \cite{Richerme2014, lerose_quasilocalized2019, rubio2020, choi2019, agrawal2022entanglement}.
 
We quantify entanglement through the bipartite mutual information between two halves of the spin chain, which isolates scrambling within the chain from decoherence directly with the ancilla \cite{Touil2021}, and operator spreading through OTOCs. Building on prior characterizations of the mixed-field long-range TFIM \cite{Belyansky2020, kim2013}, we first review ballistic operator spreading in the bare TFIM as a baseline (Sec.~\ref{sec:TFIM}), then build up the ring-ancilla model term by term, with $\lambda$ alone, then $h$ and $h_A$ (Sec.~\ref{sec:StarIsing}), then include $J$ and $g$, $g_A$ (Sec.~\ref{sec:RingAncilla}), using perturbative techniques to isolate the role of each coupling. For each model, we utilize exact diagonalization supplemented with Krylov-subspace time evolution \cite{Haegeman_KrylovKit_2024} to establish the full crossover across both the star- and ring-ancilla limits.

\begin{figure*}[ht]
    \includegraphics[width=0.85\linewidth]{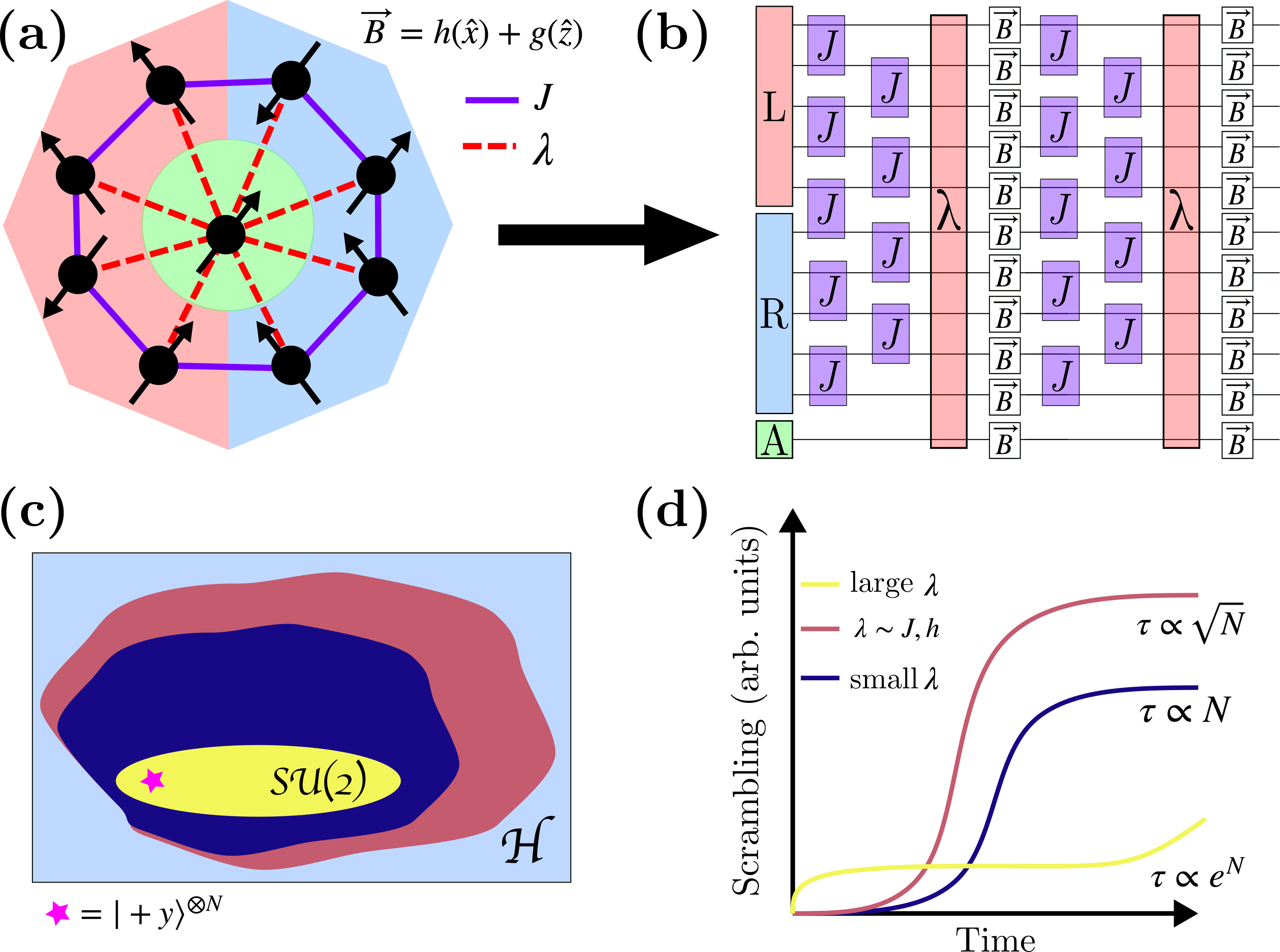}
    \caption{\label{fig:summary}
    \textbf{(a)} Schematic of the ring-ancilla system, with the left and right subsystems denoted in the blue/red background and central ancilla with green; the local Ising interaction J (solid lines) couples the left and right sites while the ancilla interaction (dashed) couples the ancilla to the local degrees of freedom of the Ising chain.
    \textbf{(b)} Schematic realization of the ring-ancilla model as a quantum circuit, with local unitary gates acting to interact the spin-1/2 qubits; such interactions may be realized through Rydberg atoms or other platforms \cite{raeisi2012quantum,seki2025simulating,liang_observation_2025,abd_rabbou_quantum_2026}.
    \textbf{(c)} Diagram of a snapshot of spread through Hilbert space $\mathcal{H}$ for an initial product state $|+y\rangle^{\otimes N}$. In the strong ancilla coupling limit, the state remains confined to the SU(2) symmetric Dicke states (yellow). In the weakly coupled limit, the state fills the entire Hilbert space allowed by the local interaction $J$ (purple), while intermediate values of $\lambda$ permit super-ballistic spreading to a greater volume of Hilbert space (red) through effective ancilla-mediated long-range interactions.
    \textbf{(d)} Scrambling in the ring-ancilla model falls into three categories. For small $\lambda$, the system scrambles in a time proportional to system size $N$ due to ballistic spreading through local interactions. When $\lambda \sim J,h$, the effective all-to-all coupling through the ancilla allows faster spreading and scrambling times $\propto \sqrt{N}$. When $\lambda$ becomes very large, it confines the state into the SU(2) Dicke manifold for times $\propto N$, only realizing full thermalization for times exponential in the system size.
    }
\end{figure*}

\subsection{Entanglement \& Scrambling Metrics}
We consider the spread of operators in Hilbert space through the conventional von Neumann entropy, defined as the entropy of the reduced density matrix (RDM) of a particular subsystem $A$:
\begin{equation}
\label{eq:vn_entropy_def}
S_{vN} = -\sum_k \lambda_k \log \lambda_k
\end{equation}
where $\{\lambda_k\}$ are the eigenvalues of $\rho_A = \mathrm{Tr}_A [\rho]$. We consider the entropy growth $S_A$ between the ancilla and spin-chain system through the reduced density matrix of the ancilla; we also rely on the mutual information 
\begin{equation}
    \label{eq:mutual_info_def}
    I = 2S_{N/2} - S_A
\end{equation}
between two halves of the spin chain, removing the entropy contribution from the ancilla and allowing a more direct comparison between entanglement growth in the isolated spin chain and a chain with an effective bath.

In constrast, we quantify this spread of operators in space-time through OTOCs of the form 
\begin{equation}
\label{eq:otoc_full_def}
C_{VW}(i,j,t) = \langle [W_j(t),V_i(0)]^\dagger[W_j(t),V_i(0)] \rangle,
\end{equation}
which examines whether local operators on sites $i,j$ in the Heisenberg picture $V_i$ and $W_j$ commute at a time $t$. This provides a picture of how quantum information behaves in space-time and depends on the geometry of the commutivity graph of the underlying Hamiltonian. We consider the case where $V_i,W_j$ are single-site Pauli operators and evaluate the expectation value of an infinite temperature state, meaning this expression simplifies to
\begin{equation}
    \label{eq:simplified_otoc_def}
    C_{VW}(i,j,t) = 2 - 2||(W_j(t)V_i(0))^2||
\end{equation}
where $||\cdot||$ denotes the operator norm 
\begin{equation}
\label{eq:operator_norm_def}
||\hat{O}|| = \frac1{2^{N+1}}\mathrm{Tr}[\hat O].
\end{equation}

In pure, closed quantum systems, local quantum information can never be lost, but instead transforms into many-body degrees of freedom inaccessible to local measurements as the support of the operator grows subject to interaction. RDM metrics like von Neumann entropy and mutual information and operator growth metrics like OTOcs describe the same type of scrambling physics, and the scrambling time $t_{sc}$ reflects both the time for arbitrary sites to become $\mathcal{O}(1)$, as well as the time for $S_{vN}$ to become $\mathcal{O}(N)$. For OTOCs, this non-rigorous limit provides only a best-case scenario for operators traversing the system rather than providing a timescale for nontrivial operator strings to span the system (extensive operator entanglement) ~\cite{Harrow2021}. Rigorous relationships between OTOCs and Renyi$-2$ entropy have been established~\cite{Zanardi2001, Yan2020, Styliaris2021} and special cases have been studied in particular optical Hamiltonians~\cite{Garttner2017, li2017, lewis-swan2019}.

Unitary scrambling dynamics generally classifies systems into those that thermalize rapidly versus those that fail to do so. The former fast-scramblers are ergodic systems (often with variable, all-to-all or infinite range interactions) that spread information throughout the full Hilbert space in $t_{sc} \sim \log(N)$; this category includes Sachdev-Ye-Kitaev (SYK) models and non-integrable long-range Ising and XY models ~\cite{KitaevKITP, Iyoda_2018, Li2020, Belyansky2020, Bentsen_sparse2019}. Slow-scrambling models that thermalize in $t_{sc}\sim e^N$ fail to obey the Eigenstate Thermalization Hypothesis (ETH) and as such are candidates for coherent quantum information storage \cite{nandkishore_many-body_2015}. This non-ETH physics arises under a variety of conditions, including integrability \cite{Rigol2008}, disorder-free localization~\cite{Hart2021, jeyaretnam_hilbert_2025}, quantum scarring ~\cite{turner2018, choi2019, serbyn2021, chandran2023quantum}, and higher order exact or proximate conservation laws~\cite{prem2017, pai2020, feng2022}. 

Recent quantum simulation results have studied collective spin models such as the Lipkin-Meshkov-Glick (LMG) model or Dicke model, which conserve the total spin moment $\hat{S}^2$ and reduce the effective number of degrees of freedom to $\mathcal{O}(N)$, rather than $\mathcal{O}(2^N)$~\cite{latorre_entanglement_2005,lewis-swan2019, Alavirad2019, lerose2020}. These systems have been observed to spread information rapidly, while the complexity saturation value remains low, in stark contrast to fast-scramblers like the SYK model, where infinite-range connectivity allows for rapid and complex quantum information scrambling. This raises the puzzle of how long-range interactions, tending toward generating semiclassical behavior, compete with local chaotic quantum dynamics to allow a fast-to-slow scrambling transition. 

A complete understanding of quantum information physics hinges on understanding both the unitary dynamics and non-Hermitian processes present in open quantum systems and inherent to quantum simulation platforms. More generalized quantum dynamical behavior has been explored in recent studies consisting of non-Hermitian operations: composite system-environment undergoing quantum measurement~\cite{Li2019, Skinner2019, Jian2020, Bao2020, Lavasani2021, Minato2022}, light-matter interactions~\cite{lewis-swan2019}, dissipative and driven systems~\cite{luschen2017, Choi2020, lenarvcivc2020, wybo2020}. Surprisingly, these works find that non-unitary dynamics generate effective inter-system interactions and impose effective static long-lived symmetries; for example, periodically driven Floquet systems resemble various unitary scrambling phases \cite{rubio2020,peng2021,potirniche2017,yin2021,sahu2025information}. For this reason, we also study the dynamics of the spin-chain after integrating out the ancilla qubit to observe how effective non-Hermitian dynamics can emerge from unitary interactions wherein the ancilla acts either as a weak probe of the local spin-chain dynamics or as a strong drive to the collective spin-chain. Inverting the perspective, we also gain insight to how the spin chain behaves as a structured bath for the two-level ancilla.

\section{\label{sec:TFIM} Transverse Ising Chain}

\begin{figure*}
    \includegraphics[width=0.85\linewidth]{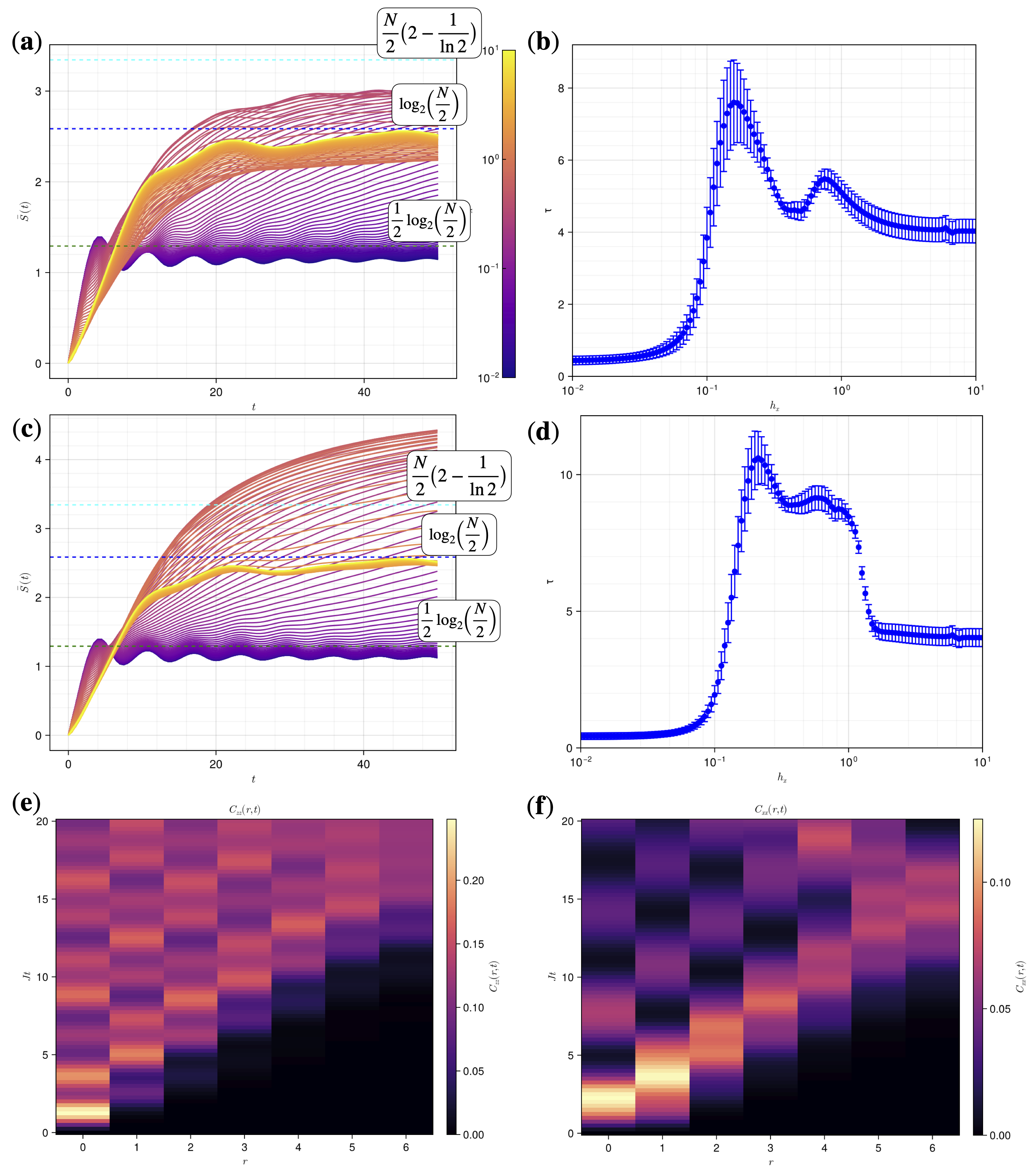}
    \caption{\label{fig:tfim_entropy_baseline}
    von Neumann entropy and OTOCs for a bare TFIM chain in isolation from the ancilla. \textbf{(a)} Average entropy after a quench from $\ket{+y}^{\otimes N}$ for $N=12, J=1, g=0, h\in[10^{-2},10]$; in the extreme limits $h/J\rightarrow 0,\infty$, the entropy saturates to a prethermal plateau determined by the underlying symmetry of the effective Hamiltonian, before eventually thermalizing to the full maximum entropy of the thermal ensemble dictated by the initial state. 
    \textbf{(b)} The extracted saturation time $\tau$ for each trace in \textbf{(a)}, plotted vs. $h$.
    \textbf{(c)} Average entropy after a quench from $\ket{+y}^{\otimes N}$ for $N=12, J=1, g=0.1, h\in[10^{-2},10]$; here we see the same prethermal behavior in the limiting cases, but the intermediate regime thermalizes well beyond the limit for the integrable model; the longitudinal field term mixes fermion occupations, allowing the system to explore the entire Hilbert space. 
    \textbf{(d)} the extracted saturation time $\tau$ for each trace in \textbf{(c)} plotted vs. $h$; note the saturation times are larger across the full parameter regime, reflecting the larger available Hilbert space of the nonintegrable model.
    \textbf{(e), (f)} $C_{zz}(r,t)$ and $C_{xx}(r,t)$ for $N=12, J=1,g=0.1,h=1.25$ showing the ballistic spreading of operators, following the Lieb-Robinson light cone.
    }
\end{figure*}

We begin by considering the mixed field Ising chain Hamiltonian
\begin{equation}
    \label{eq:bare_tfim_hamiltonian}
    \mathsf{H_{TFIM}} = J\sum_j S_j^z S_{j+1}^z - h S^x - g S^z
\end{equation}
when $g=0$, this model has a well-understood critical point at $h/J=1/2$ and well-defined Bogoliubov fermion quasiparticle excitations following the dispersion \cite{pfeuty1970one, sachdev2011qptbook,mbeng_quantum_2020}
\begin{equation}
    \label{eq:tfim_dispersion}
    \varepsilon_k = \frac{J}{2}\sqrt{1 + (2h/J)^2 - (4h/J)\cos k}.
\end{equation}

When $g=0$, the local Ising chain exhibits ballistic entanglement growth up to a saturation time $\tau \propto N/2v_{LR}$ in terms of the Lieb-Robinson velocity $v_{LR}$ associated with the fastest dispersing quasiparticle modes (the factor of two results from periodic boundaries). This maximal velocity, independent of system size, follows from the dispersion Eq.~\eqref{eq:tfim_dispersion} near the critical point, which provides $v_{LR}\leq J/2$ and an upper bound for the saturation time $N/J$. The maximum entropy is determined by that of the conserved density of Bogoliubov fermions $\langle n_k\rangle$. 

We consider the zero-energy product state $\ket{+y}^{\otimes N}$ as an effective infinite temperature state with $\langle \mathsf{H}\rangle = 0$; this state has previously been used to characterize high-energy dynamics of mixed-field Ising chains \cite{Belyansky2020}; in particular, this state's local observables rapidly converge to the expected thermal values, a phenomeon labeled ``strong thermalization'' \cite{banuls_strong_2011}. The fermion density in this state evaluates to
\begin{equation}
\label{eq:fermion_density_plus_y}
\langle n_k\rangle = \frac1{2}\Big[1-\frac{2h/J\cos k - \cos 2k}{\sqrt{1 + (2h/J)^2 - (4h/J)\cos k}}\Big]
\end{equation}
from which we determine the thermal entropy per site
\begin{equation}
\label{eq:yang-yang_entropy}
s_{th} = \frac1{\pi}\int_0^\pi dk\ [-n_k\log_2(n_k)-(1-n_k)\log_2(1-n_k)]
\end{equation}
which evaluates numerically to $s_{th}\lesssim s_0\equiv 2-\frac1{\ln(2)}<1$ (see details in Supplemental Material). 

The entropy for this quench computed numerically with exact diagonalization is shown in Fig. \ref{fig:tfim_entropy_baseline}(a) for a range of $h$ values. We characterize this by the cumulative average half-chain entropy
\begin{equation}
\bar{S}(t) \equiv \frac1{t} \int_0^t ds S_{vN}(s)
\end{equation}
and fit this function to an average exponential form 
\begin{equation}
\label{eq:average_exp_fit}
\bar{S}(t) = S_{\infty} \Big(1-\frac{1-e^{-t/\tau}}{t/\tau}\Big)
\end{equation}
to extract the saturation time scale $\tau$ and saturation entropy $S_\infty$. In the limiting cases, we see the entropy reach a stable prethermal plateau value, from which we extract $\tau$, before fully thermalizing to the bound $s_0$ only for very long times exceeding our numerical data. In the large field limit $h\gg J$, we treat the local coupling $J$ as a perturbation and perform a Schrieffer-Wolff (SW) transformation (detailed in the Supplemental Material) that removes Hamiltonian terms that fail to commute with $S^x$; thus, the entropy saturates to $\log_2(\frac{N}{2})$, reflecting that the conserved magnetization traps the state in a particular symmetry sector whose Hilbert space grows linearly in $N$. In the small field limit $h\ll J$, the effective Hamiltonian conserves $S^z$ and the domain wall number $N_d = \sum_j \frac1{2}(1-\sigma_j^z\sigma_{j+1}^z$); this prevents hard-core domain walls from moving through one another, constraining the allowed motion of quasiparticles. The conservation of these two U(1) symmetries combined results in a further restriction of the Hilbert space like $\sqrt{N}$ \cite{sala_ergodicity_2020,yang_hilbertspace_2020,royen_enhanced_2024}. Implementing the longitudinal field $g\neq0$ destroys the integrable nature of the model and mixes $\langle n_k\rangle$ sectors; without the extensive conserved quantities of the integrable model, the system is free to thermalize to the full Page limit of $N$ bits, while still subject to the prethermal plateaus near the integrable limits.

In the real space-time operator picture, we can write $S_j^z$ at early times using the Baker-Campbell-Hausdorff (BCH) expansion:
\begin{equation}
\hat O(t) = e^{i\mathsf{H}t}\hat Oe^{-i\mathsf{H}t} = \sum_{m=0}^\infty \frac{(it)^m}{m!} [\mathsf{H},\hat O]_m
\end{equation}
\begin{equation}
[A,B]_m = [A,[A,B]]_{m-1}; [A,B]_0 = B
\end{equation}
from which we have
\begin{align}
S_i^z(t) \approx (1&- \frac{(ht)^2}{2})S_i^z + ht S_i^y + \frac{ght^2}{2} S_i^x \nonumber\\ &+ \frac{hJt^2}{2}S_i^x(S_{i-1}^z + S_{i+1}^z) +\mathcal{O}(t^3).
\end{align}
From this expression, we observe from the trend in operator growth that
\begin{equation}
||S_i^z(t) S_j^z|| \propto \mathrm{Tr}[S_i^z(t) S_j^z] \sim \frac{t^{|i-j|}}{|i-j|!} ,
\end{equation}
and the OTOC scales like the square of this:
\begin{equation}
    \label{eq:tfim_otoc_scaling}
    C_{zz}(i,j,t) \sim \frac{t^{2|i-j|}}{(|i-j|!)^2}
\end{equation}
for times $t\sim |i-j|/v_B$, where $v_B$ is a characteristic butterfly velocity \cite{roberts2016}. Before this time, operator growth is suppressed with an exponent that grows with $r = |i-j|$, and allows for optimized simulations using matrix product operator dynamics (MPO) by just keeping track of the operator wavefront~\cite{lieb_finite_1972,hastings_locality_2010}. This exponentially growing operator weight leads to the development of a linear light cone with butterfly velocity exactly calculated as $eJ$ ($e$ being Euler's number) for $2h/J > 1$~\cite{Chen_2021}. Following this light cone in the nonintegrable Ising model, any localized operator spans the full operator Hilbert space $\{\sigma^x, \sigma^y,\sigma^z, \mathbb{I}\}$ within region of radius $r$. We can also consider the integrated OTOC as a measure of the complexity growing within the radius $r$ \cite{Keselman2021}. In the nonintegrable Ising model, an operator saturates the reduced Hilbert space $4^{|i-j|}$ after a time $t_{sat} = \frac{|i-j|}{v_B}$, so integrating Eq.~\eqref{eq:tfim_otoc_scaling} over $r=|i-j|$ gives an exponentially growing iOTOC for all $C_{VW}$. This operator growth then guarantees ballistic growth of entanglement entropy, as $S_{vN} \sim \sum_{V,W}\log[\text{iOTOC}(V,W)]$ for all single-site operators $V,W$. This butterfly velocity is demonstrated in Fig.\ref{fig:tfim_entropy_baseline}(c),(d) for the OTOCs $C_{zz}(r,t),C_{xx}(r,t)$, showing the local operator spreading over many sites, following the Lieb-Robinson light cone.

\section{\label{sec:StarIsing} Star Ising Model}

Next consider the so-called star-Ising model given by Eq.~\eqref{eq:full_ring_star_ham} with $J=0$. We first consider the limit with no fields $h=h_A=g=g_A=0$; this system is exactly solved in the $S^z$-basis, since $[S_j^z,\mathsf{H}]=0\forall j,A$. Utilizing the BCH expansion for 
\begin{equation}
S_j^x(t) = S_j^x \cos(\frac{\lambda}{2} t) - 2S_j^y S_A^z \sin(\frac{\lambda}{2} t)
\end{equation}
one can show the autocorrelation
\begin{equation}
||S_j^x(t)S_j^x|| = \frac1{2}\mathrm{Tr}[\hat S_j^x(t)S_j^x] = \frac1{4}\cos(\frac{\lambda}{2}  t)
\end{equation}
and the OTOC 
\begin{equation}
    \label{eq:coherent_autocorr_result}
    C_{xx}(i,j,t) = \frac1{4}\sin^2(\frac{\lambda}{2} t)\delta_{i,j}.
\end{equation}
This seemingly trivial result provides key insight to how operators orthogonal to the ancilla interaction initially propagate to the ancilla in time $\pi/\lambda$ and coherently oscillate between the chain site and the ancilla. An orthogonal operator initialized on the ancilla oscillates onto every site not as a complex many-body operator, but as a collective superposition of two-body operators. While the operators become rapidly nonlocal, minimal entanglement entropy develops since the number of symmetric operator states scales $\sim N$ in a Hilbert space of size $4^N$; quench experiments on such collective spin systems restrict entanglement entropy to a maximum of $\sim\log N$ \cite{szabo2022}. 

Reinstating only the transverse field on the ancilla $h_A\neq0$, the BCH expansion acquires additional terms
\begin{align}
    \label{eq:Sx_BCH_extra-terms}
    S_j^x(t) = &S_j^x \cos(\frac{\lambda}{2} t) - 2S_j^y S_A^z \sin(\frac{\lambda}{2} t) + \frac{\lambda h_A t^2}{2} S_j^y S_A^y \nonumber\\ &+ \frac{\lambda h_A^2t^3}{6}S_j^y S_A^z - \frac{\lambda^2 h_A t^3}{6}S_j^y S_A^x \sum_{i\neq j}S_i^z + \mathcal{O}(t^4)
\end{align}
where the non-commuting transverse field now enables operator weight to grow through an effective coupling directly between all chain sites (more detail in Supplemental Material). The system remains integrable in the $S^z$-basis for the spin-chain $\ket{z}$ (where $z\in \mathbb{Z}$ maps to the $N$-digit bitstring whose $j$-th digit corresponds to the eigenvalue of $(1/2-S_j^z)$). In each sector with $m_z = \sum_{j=1}^N S_j^z$ sector, the ancilla experiences a magnetic field along the direction $h_A(\hat{x}) -\lambda m_z(\hat{z})$, so the unitary time evolution acting on a basis state is
\begin{align}
    \label{eq:star_ising_basis_time_evolution}
    e^{-i\mathsf{H}t} \ket{z}\otimes&\ket{\pm z}_A = \Big[\cos(\omega_{z}t) \nonumber\\ &-i \sin(\omega_{z}t)\frac{\lambda m_zS_A^z-h_A S_A^x}{\omega_{z}}\Big]\ket{z}\otimes \ket{\pm z}_A
\end{align}
where we have used the Euler identity $e^{i(\vec{v}\cdot\vec{\sigma})t} = \cos(|v|t) + i\sin(|v|t) \frac{\vec{v}\cdot \vec{\sigma}}{|v|}$ and defined the field magnitude 
\begin{equation}
    \label{eq:ancilla_star_field}
    \omega_{z} = \frac1{2}\sqrt{(\lambda m_z)^2+h_A^2}.
\end{equation}
The operator $S_j^x$ flips a single spin in the state $\ket{z}$, meaning that the new state will subject the ancilla to the Hamiltonian with $\tilde m_z$ = $m_z-(-1)^{[z]_j}$ (with $[z]_j$ denoting the $j$-th binary digit of $\ket{z}$). Altogether, the operator dynamics are given by 
\begin{align}
\label{eq:autocorr_full_sum_form}
4||S_j^x(t)S_j^x|| = &\frac1{2^{N+1}}\sum_{\pm z_A}\sum_{\ket{z}} \cos(\omega_{z}t)\cos(\tilde \omega_{z}t) \nonumber\\ &+ \frac{\omega_z^2-\lambda^2 m_z(-1)^{[z]_j}}{\omega_{z} \tilde \omega_{z}} \sin(\omega_{z} t) \sin(\tilde \omega_{z}t).
\end{align}
To evaluate this expression, note that the sum over states $\ket{z}$ can be recast as a sum over $m_z$ sectors, weighted by the binomial coefficient $\binom{N}{N/2-m_z}$; in the thermodynamic limit $N\rightarrow \infty$, this becomes an integral over a Gaussian distributed variable $m_z$ with mean $0$ and variance $N/4$, so we change to the variable $\bar m_z \equiv 2m_z/\sqrt{N}$ and evaluate the sum as an integral over the Gaussian kernel
\begin{equation}
K(\bar m_z) =\frac{e^{-\bar m_z^2/2}}{\sqrt{2\pi}}.
\end{equation}
This also means that the behavior is dominated by low-frequency dephasing of states with $|m_z|\ll N/2$, since they dominate the Hilbert space. 

In the weak coupling limit $\lambda \ll h_A$, we fix the ratio $\alpha = h_A/N$ and Taylor expand the magnitude of the ancilla field Eq.~\eqref{eq:ancilla_star_field} as 
\begin{equation}
\omega_z \approx \frac{\alpha N}{2}\Big[1+\frac1{2}\Big(\frac{\lambda \bar m_z}{2 \alpha \sqrt{N}}\Big)^2\Big]+\mathcal{O}(1/N).
\end{equation}
From this we see that the coefficient of the sine term in Eq.~\eqref{eq:autocorr_full_sum_form} is $\mathcal{O}(1)$; using the identity $\cos(A)\cos(B) = \cos(A-B) - \sin(A)\sin(B)$, the $\sin$ terms vanish in the thermodynamic limit, leaving dynamics driven by transitions between states with $m_z$
and $m_z\pm 1$ with a dominant frequency of
\begin{equation}
\Delta\omega_z = \omega_z - \tilde \omega_{z}\approx \frac{\lambda^2}{4h_A}(1\pm \bar m_z \sqrt{N});
\end{equation}
evaluating the integral of $\cos(\Delta\omega_zt)$ over the Gaussian kernel yields 
\begin{equation}
    \label{eq:weak_coupling_star_operator_growth}
||S_j^x(t)S_j^x||_{\text{weak-}\lambda} = \frac1{4}\cos\Big(\frac{\lambda^2}{4h_A} t\Big)\exp\Big[-\frac1{2}\Big(\frac{\lambda^2\sqrt{N}}{4h_A} t\Big)^2\Big]
\end{equation}
Here the strong applied field leads to rapid decoherence on a time scale $\lambda\tau\sim \frac{4h_A}{\lambda\sqrt{N}}$ along with a slow oscillating modulation. Including the longitudinal field $g_A\neq0$ modifies the effective ancilla field $\omega_z$, replacing $-\lambda m_z \rightarrow g_A-\lambda m_z$; the full details are shown in the Supplementary Material.

Taking the strong-coupling limit $\lambda \gg h_A$, we instead fix the ratio $\beta \equiv h_A/\sqrt{N}$, rewriting the ancilla field as
\begin{equation}
\omega_z = \frac{\sqrt{N}}{2} \sqrt{\beta^2 + \Big(\frac{\lambda \bar m_z}{2}\Big)^2}.
\end{equation}
The dominant dephasing frequency is given by  
\begin{equation}
\Delta \omega_z \approx \frac{d\omega_z}{d \bar m_z}(\pm1) = \pm \frac{\lambda^2 \bar m_z}{2\sqrt{(\lambda\bar m_z)^2+4\beta^2}}
\end{equation}
which yields the following integral equation for the weak-field dynamics:
\begin{align}
    \label{eq:low_field_star_operator_growth_integral}
    ||S_j^x(t)S_j^x|| = &\frac1{4}\int_{-\infty}^\infty \frac{d\bar m_z}{\sqrt{2\pi}} e^{-\bar m_z^2/2} \nonumber\\ &\times \cos\Big(\frac{\lambda t}{\sqrt{1+(2\beta/
    \lambda \bar m_z)^2}}\Big).
\end{align}
Expanding the denominator in the limit $\lambda \bar m_z \gg \beta$ gives 
\begin{equation}
\cos(\Delta \omega_zt) \approx \cos\Big(\frac{\lambda}{2} t\big[1-\frac1{2}\big(\frac{2\beta}{\lambda \bar m_z}\big)^2\big]\Big)
\end{equation}
which gives a closed form solution:
\begin{align}
    \label{eq:strong_coupling_star_operator_growth_solution}
    ||S_j^x(t)S_j^x||_{\text{strong-}\lambda}\approx &\frac1{4}\cos\Big(\frac{\lambda}{2} t-\frac{h_A}{\sqrt{\lambda N}}\sqrt{t}\Big)\nonumber\\ &\times \exp\Big[-\frac{h_A}{\sqrt{\lambda N}}\sqrt{t}\Big].
\end{align}
For small fields, we recover rapidly oscillating dynamics, approaching the coherent result $\cos(\frac{\lambda}{2} t)$ as $h_A\rightarrow 0$, with a slow decay on the time scale $\lambda \tau \sim N\lambda^2/h_A^2$. 
Expanding the denominator instead with $\lambda\bar m_z \lesssim\beta$, the integrand becomes 
\begin{equation}
\cos(\Delta \omega_zt)\approx\cos\Big( \frac{\lambda|\bar m_z|}{4\beta}\lambda t\Big)
\end{equation}
in the relevant window $|\bar m_z| < \frac{\beta}{\lambda}$, which provides a small $\mathcal{O}(\beta/\lambda)$ correction to the low-field expansion result Eq.~\eqref{eq:strong_coupling_star_operator_growth_solution}. Including the longitudinal field does not significantly modify the results in this regime, as terms $\propto g_A/\sqrt{N}$ vanish in the thermodynamic limit.

\begin{figure}
    \includegraphics[width=0.85\linewidth]{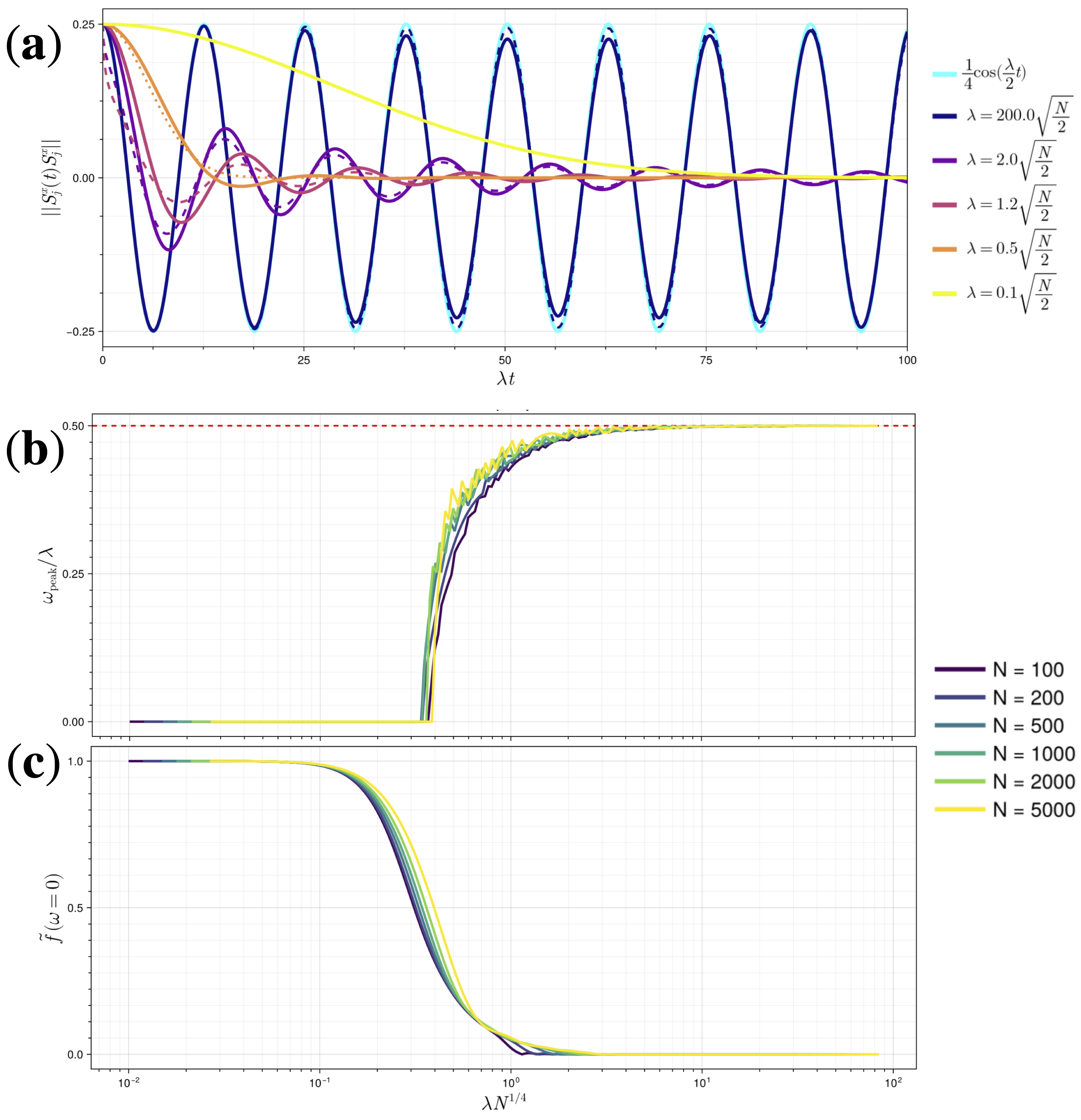}
    \caption{\label{fig:autocorr}
    Infinite temperature autocorrelation of $S_j^x$.
    \textbf{(a)} Plot of Eq.~\eqref{eq:autocorr_full_sum_form} vs $\lambda t$ for $N=1000, h_A=1$ and varying $\lambda$ in multiples of $\sqrt{N/2}$; the dotted and dashed lines follow the respective weak- and strong-coupling limiting forms in Eqs.~\eqref{eq:weak_coupling_star_operator_growth} and ~\eqref{eq:strong_coupling_star_operator_growth_solution}. The coherent, $h_A=0$ form Eq.~\eqref{eq:coherent_autocorr_result} is shown in cyan, approximately approached by the trace for $\lambda = 200\sqrt{N/2}$.
    \textbf{(b)} Dominant peak frequency in the Fourier transform of Eq.~\eqref{eq:autocorr_full_sum_form} (in units of $\lambda$) vs. $\lambda N^{1/4}$ for varying system sizes $N$. For large $\lambda$ the system approaches the coherent peak at $\omega/\lambda = 1/2$, while for small $\lambda$ the system is dominated by the decaying dynamics centered on $\omega=0$.
    \textbf{(c)} Amplitude of the DC component of the Fourier transform of Eq.~\eqref{eq:autocorr_full_sum_form} vs. $\lambda N^{1/4}$ for varying system sizes $N$, again showing the prevalence of $\omega=0$ at small $\lambda$.
    }
\end{figure}

The dynamics of the system are dominated by the longer of the time scales in the two limits, from which we identify a crossover at field strength $h_A \sim\lambda \sqrt{N}$. As the field increases, greater operator weight $S_A^z$ decoheres more rapidly into $S_A^{x,y}$; these operators fail to commute with the Ising interaction and transfer weight onto sites $i\neq j$, leading to decay in the autocorrelation. Analyzing the Fourier spectrum of the autocorrelation in each limit, we see that the strong-coupling case is dominated by a peak at $\omega=\lambda/2$, approaching the coherent result Eq.~\eqref{eq:coherent_autocorr_result}, while the weak-coupling case has a dominant frequency at $\omega=\lambda^2/4h_A$ (which scales $\sim\mathcal{O}(1/N)$ in this limit). Decreasing the coupling strength dampens the oscillations until eventually, the relatively large magnetic field quickly rotates operators $S_A^z$ into superpositions of two-body operators with weight $1/N$. This behavior is shown in Fig.~\ref{fig:autocorr} with both finite $N$ and the asymptotic, large-$N$ forms. The transition between regimes depends on the width of the zero-frequency Fourier peak masking the peak near $\lambda/2$; as the ancilla coupling increases, the large DC peak becomes narrower, eventually revealing and being overtaken by the growing  $\lambda/2$ oscillation. Both the zero-frequency peak amplitude and dominant Fourier frequency scale like $\lambda N^{\frac1{4}}$ due to the scaling of the Gaussian width $\tau \propto 1/\lambda^2 \sqrt{N}$ in the weak-coupling limit.

\subsection{Weak Ancilla Coupling Limit}

We gain further insight to the entanglement growth within the collective spin system by defining an effective Hamiltonian for the isolated chain in a particular ancilla basis using a Feshbach-Fano projection, formally described in the Supplemental Material. Considering the model with only ancilla coupling $\lambda$ and transverse fields $h,h_A$ and taking the weak-coupling limit $\lambda\ll h,h_A$, we define $\mathsf{P_x} = \ket{+x}\bra{+x}_A$, the orthogonal projector $\mathsf{Q_x} = \mathsf{I}-\mathsf{P_x} = \ket{-x}\bra{-x}_A$, and the projected Hamiltonian
\begin{equation}
\mathsf{P_x HP_x} = -hS^x -h_A/2.
\end{equation}
In this basis, the spin chain system is diagonal in the $x$-magnetization basis $\ket{S,m_x}$, and the ancilla coupling term acts as a perturbation mapping an initial state $\ket{S,m_x}\ket{+x}_A$ with energy $E_0 = -hm_x-h_A/2$ onto the state $\ket{S,m_x\pm1}\ket{-x}_A$ with the energy $-hm_x\mp h +h_A/2$, thus giving an energy difference $E_0 - \mathsf{Q_x HQ_x} = h_A\pm h$ and the effective Hamiltonian
\begin{equation}
\mathsf{H_{eff}^{P_x}} = -hS^x -h_A/2 +\frac{\lambda^2}{4}(S^z)^2 \Big(\frac1{h_A-h}+\frac1{h_A+h}\Big).
\end{equation}
This Hamiltonian closely resembles the Lipkin-Meshkov-Glick model, which hosts a well understood dynamical phase transition when $\lambda^2 \sim \frac{h_A h}{N}$. The weak coupling regime describes a dynamical paramagnet, wherein the long-time average of $S^z$ is always 0, no matter the initial state; in this limit, entanglement grows with a rate $\sim \lambda^2$ as states with differing $m_x$ dephase at different rates while the longitudinal interaction creates one-axis twisting. The strong-coupling regime describes a dynamical ferromagnet where $S^z$ retains a nonzero finite expectation value dependent on the initial state, since the easy-axis term $(S^z)^2$ dominates \cite{bento2024krylov}. 

We have a Feshbach resonance when $h=\pm h_A$ which results from the degeneracy between the states $\ket{S,m_x}\ket{+x}$ and $\ket{S,m_x\pm1}\ket{-x}$; the perturbation theory breaks down in this degenerate subspace, which is governed by the degenerate two-state Hamiltonian
\begin{equation}
\mathsf{H_{eff}}(m_x)=
\begin{pmatrix}
    -h(m_x\pm\frac1{2}) & \frac{\lambda}{2}\eta(m_x) \\ \frac{\lambda}{2}\eta(m_x) & -h(m_x\pm\frac1{2}).
\end{pmatrix};
\end{equation}
\begin{equation}
     \eta(m_x)= \sqrt{s(s-1)-m_x(m_x\pm1)}.
\end{equation}

\subsection{Strong Ancilla Coupling Limit}

If we instead consider the strong coupling limit $\lambda \gg h_A$, we use the projectors $\mathsf{P_z} = \ket{+z}\bra{+z}_A, \mathsf{Q_z} = \ket{-z}\bra{-z}_A$ and identify the transverse field term $-h_AS_A^x$ as the perturbation. The projected Hamiltonians formed by each of these operators are
\begin{equation}
\mathsf{P_z HP_z} = \frac{\lambda}{2}S^z - hS^x;
\end{equation}
\begin{equation}
\mathsf{Q_z HQ_z} = -\frac{\lambda}{2}S^z - hS^x.
\end{equation}
$\mathsf{P_z HP_z}$ is solved in a basis rotated along the direction $h(+\hat x) + \frac{\lambda}{2}(\hat z)$ with $B^2 \equiv h^2 + \lambda^2/4$:
\begin{equation}
\begin{cases}
    S^z &= -\frac{\lambda}{2B}\tilde S^z -\frac{h}{B}\tilde S^x \\
    S^x &= \frac{h}{B}\tilde S^z - \frac{\lambda}{2B}\tilde S^x
\end{cases}
\end{equation}
which gives the rotated operators
\begin{equation}
\mathsf{P_z\tilde H P_z} = -B \tilde S^z
\end{equation}
\begin{equation}
\mathsf{Q_z \tilde H Q_z} = \frac{B^2 - 2h^2}{B}\tilde S^z + \frac{h\lambda}{B}\tilde S^x
\end{equation}
$\mathsf{P_z\tilde H P_z}$ is diagonal in the magnetization basis $\ket{S,\tilde m_z}$ with energies $-B \tilde m_z$; the perturbation flips the ancilla spin without modifying the chain state, but subjects the chain to the Hamiltonian $\mathsf{Q_z\tilde H Q_z}$ for its duration in this sector, during which the state vector decoheres under the action of $\tilde S^x$. The energy gap between sectors is $\frac{\lambda^2}{2B}\tilde m_z$, yielding the effective Hamiltonian 
\begin{equation}
\mathsf{\tilde H_{Eff}^{P_z}} = -B \tilde S^z - \frac{Bh_A^2}{2\lambda^2}\frac1{\tilde S^z}.
\end{equation}
When $|\tilde m^z|$ is large, the $1/\tilde S^z$ term provides a linear confining potential. Instead of $S^z(t)$ precessing as in the weak-coupling limit, the wave function amplitudes in each $m_z$ sector become frozen in their initial configuration, and we observe constant behavior for extended times in the expectation value and variance of $S^z(t)$ (see the Supplemental Material for details). Clearly, this perturbation theory breaks down in the presence of the Feshbach resonance in the $\tilde m_z=0$ sector, where the ancilla states become degenerate, allowing entanglement to grow between the $\mathsf{P_z,Q_z}$ subspaces.

We further develop the behavior of the $1/\tilde S^z$ effective potential in the strong coupling limit $\lambda \gg h,h_A$ by block-diagonalizing the Star-Ising Hamiltonian in terms of eigenvalues $m_z$ of the total moment of the spin sites $S^z$. We define the block diagonal Hamiltonian
\begin{equation}
    \mathsf{H_0}(m_z) = \lambda m_zS^z_A - h_A S^x_A,
\end{equation}
and the local noncommuting perturbation
\begin{equation}
    \mathsf{V} = -h S^x.
\end{equation}
Within each $m_z$ sector, the ancilla experiences an effective Zeeman field with magnitude $B(m_z) = \sqrt{h_A^2 + (\lambda m_z)^2}$ with the ancilla energy eigenvalues $\epsilon_\pm(m_z) = \pm\frac1{2}B(m_z)$ and ground and excited states:
\begin{equation}
\ket{g(m_z)} = \sin(\tfrac{\theta_{m_z}}{2})\ket{0}+\cos(\tfrac{\theta_{m_z}}{2})\ket{1},
\end{equation}
\begin{equation}
\ket{e(m_z)} = \cos(\tfrac{\theta_{m_z}}{2})\ket{0}-\sin(\tfrac{\theta_{m_z}}{2})\ket{1}
\end{equation}
in terms of the azimuthal Bloch angle 
\begin{equation}
    \tan\theta_{m_z} \equiv (h_A/\lambda m_z);\ \theta_0 = \pi/2.
\end{equation}
We define the projectors onto the ground and excited state spaces of each sector $\mathsf{P}_{g,e}$:
\begin{equation}
\mathsf P_{g}(m_z) = \Pi_{m_z}\otimes|g(m_z)\rangle\langle g(m_z)|;
\end{equation}
\begin{equation}
\mathsf P_{e}(m_z) = \Pi_{m_z}\otimes|e(m_z)\rangle\langle e(m_z)|.
\end{equation}

The overlap functions $F_{gg}^\pm$ and $F_{ge}^\pm$, which describe the projection of the ancilla ground states into neighboring sectors with $m_z\pm 1$, depend on the angle difference $\delta\theta_{m_z}^\pm \equiv \theta_{m_z}-\theta_{m_z\pm1}$
\begin{align}
F_{gg}^\pm(m_z) \equiv &\langle g(m_z)|g(m_z\pm1)\rangle = \cos(\tfrac{\delta\theta_{m_z}^\pm}{2}) \nonumber\\ &= \sqrt{\tfrac1{2}(1+\cos\delta\theta_{m_z}^\pm)}
\end{align}
\begin{align}
F_{ge}^\pm(m_z)  \equiv  &\langle g(m_z)|e(m_z\pm1)\rangle = \sin(\tfrac{\delta\theta_{m_z}^\pm}{2}) \nonumber\\&= \mathrm{sgn}(\delta\theta_{m_z}^\pm) \sqrt{\tfrac1{2}(1-\cos\delta\theta_{m_z}^\pm)}.
\end{align}
To first order, the perturbation $\mathsf{V}$ connects ground states to ground states in neighboring $m_z$ blocks; at second order, $\mathsf{V}$ connects ground to excited states. Projecting $\mathsf{V}$ onto the ground state sector, we have the first order term
\begin{equation}
    \label{eq:first-order-mz-shift}
    \mathsf{H}^{(1)}(m_z)\equiv\mathsf{P}_g\mathsf{V}\mathsf{P}_g = -\frac{h}{2} \left[ F_{gg}^+(m_z) S^+ + F_{gg}^-(m_z) S^- \right].
\end{equation}
At second order, virtual transitions project from the ground state into excited state space then back to the ground state with either the same $m_z$ or $m_z\pm 2$, giving the off-diagonal term
\begin{align}
    \label{eq:second-order-mz-shift}
    \mathsf{H}^{(2)}(m_z)&\equiv \mathsf P_g \mathsf V \mathsf P_e \frac1{B(m_z\pm 1)+B(m_z)}\mathsf P_e \mathsf V \mathsf P_g \nonumber \\ = &-\frac{h^2}{4}\Big[\frac{|F_{ge}^+(m_z)|^2}{B(m_z+1)+B(m_z)} S^- S^+ \nonumber \\ &+\frac{|F_{ge}^-(m_z)|^2}{B(m_z-1)+B(m_z)} S^+ S^-] \nonumber\\ 
    &+ \frac{F_{ge}^+(m_z)F_{ge}^+(m_z+1)}{B(m_z+1)+B(m_z)} (S^+)^2 \nonumber\\
    &+ \frac{F_{ge}^-(m_z)F_{ge}^-(m_z-1)}{B(m_z-1)+B(m_z)} (S^-)^2 \Big].
\end{align}

Near the ends of the magnetization spectrum with $|m_z|\sim\mathcal{O}(N/2)$, so long as the transverse field $h_A\ll \lambda |m_z|$, we have  $\theta_{m_z}\approx h_A/\lambda m_z$ and $\delta\theta_{m_z}^\pm\approx \mp h_A/\lambda m_z^2$. Expanding in this limit, the overlaps $F_{gg}^\pm\approx 1-\frac{h_A^2}{8\lambda^2m_z^4}$ and $F_{ge}^\pm\approx \mp\frac{h_A}{2\lambda m_z^2}$; thus, the perturbation has vanishing strength near the ends of the magnetization spectrum. In the opposing limit around $m_z=0$, we instead have $\delta\theta_{m_z}\approx \mp \lambda/h_A$, which simplifies the overlaps $F_{gg}^\pm(m_z)\approx\cos(\lambda/2h_A)$ and $F_{ge}^\pm=\mp\sin(\lambda/2h_A)$. The first-order perturbation Eq.\eqref{eq:first-order-mz-shift} suppresses the transverse field by a factor $\cos(\lambda/2h_A)$. In the second-order term Eq. \eqref{eq:second-order-mz-shift}, the terms $S^\pm S^\mp$ become $\approx \sin^2(\lambda/2h_A)/2h_A$, whereas the terms $S^\pm S^\pm$ carry an additional minus sign ($\mathrm{sgn}(\delta\theta_{m_z}^\pm)$)from the transition back from excited state space. Factoring out the amplitudes and simplifying the remaining operators $S^-S^+ + S^+ S^- - (S^+)^2 - (S^-)^2 = 4(S^y)^2$ reveals the effective all-to-all coupling:
\begin{equation}
    \mathsf{H}^{(2)}(m_z) \approx -\frac{h^2}{h_A} \sin^2(\lambda/2h_A) (S^y)^2
\end{equation}

Each $S^z$ component of the state periodically dephases, only able to decohere across sectors at very long times. The expansion near $m_z\sim 0$ breaks down as $\lambda\gg h_A$, where the perturbation energy scale overtakes the energy gap to flip the ancilla $\sim h_A$. This provides the mechanism for prethermal oscillations with an amplitude limited by the information capacity of the ancilla, observed at early times in the mutual information in Figure \ref{fig:star-ising-scaling}(a), as the perturbation allows the ancilla to tunnel between nearly degenerate states.

\subsection{Mutual Information Dynamics}

\begin{figure*}
    \includegraphics[width=0.85\linewidth]{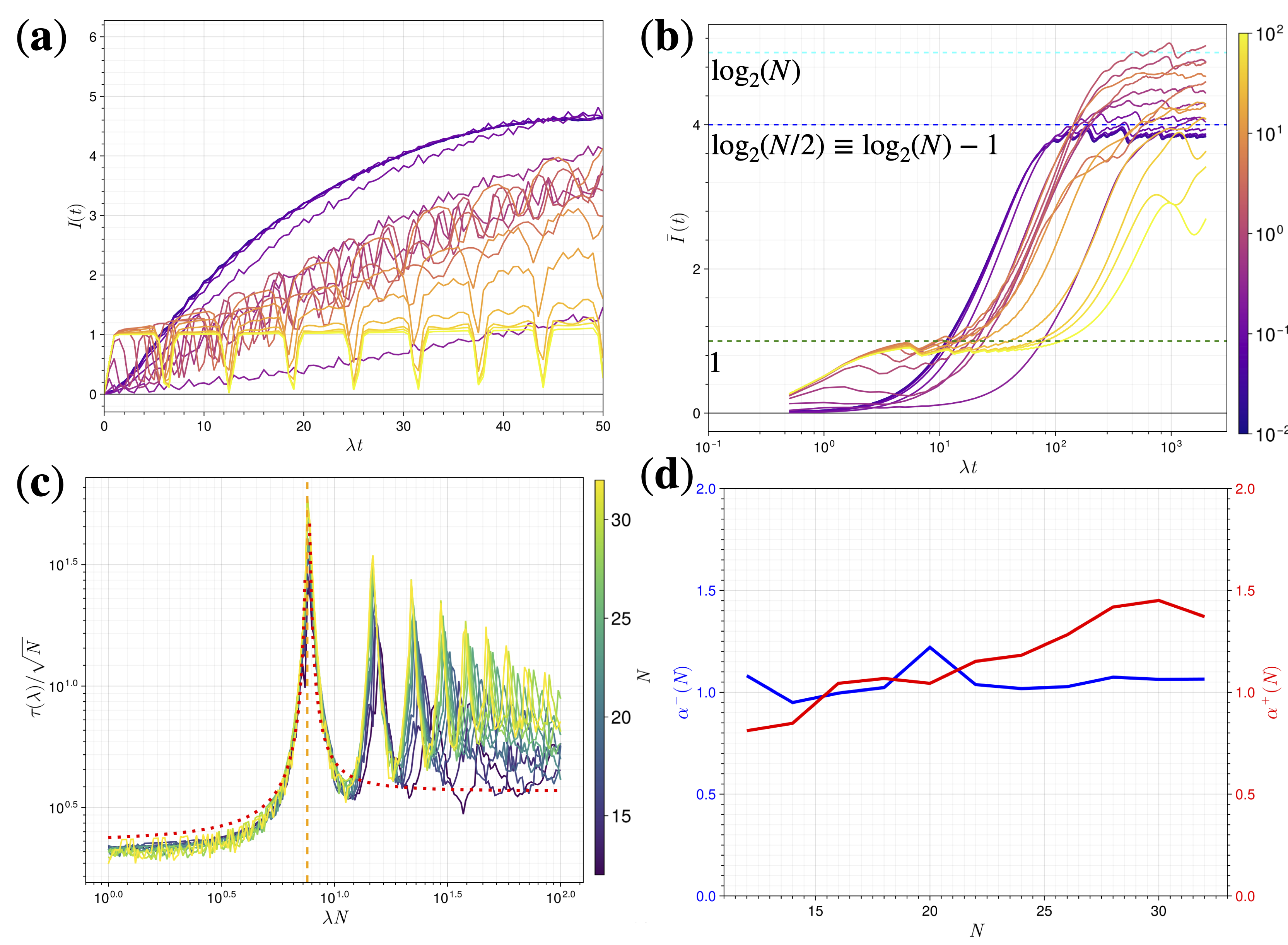}
    \caption{\label{fig:star-ising-scaling}
    Entanglement growth results for the Star-Ising model with $J=0,g=g_A=0$. 
    \textbf{(a)} Half-chain mutual information $I(\lambda t)$ for $N=32, \lambda \in [10^{-2},10^2], h=h_A=1.05$.  
    \textbf{(b)} Average half-chain mutual information $\bar{I}(\lambda t)$ for $N=32, \lambda \in [10^{-2},10^2], h=h_A=1.05$; for $N\lambda \ll h$, $\bar I(t)$ saturates to $I_0\equiv \log_2(N)-1$, while for $N\lambda\gg h$, it quickly saturates to $1$, before slowly fully thermalizing to $I_0$. In the critical region, the ancilla enhances entanglement to saturate to a maximum of $I_\mathrm{max}\equiv \log_2(N)$. 
    \textbf{(c)} Fitting for the early-time saturation time scale $\tau$ vs. $N\lambda$ for even $N\in[12,32]$, we identify the dynamic transition by the divergence (sharp maximum) in the saturation time scale, indicating critical slowing. This transition occurs for fixed $N\lambda_c \approx 8$. 
    \textbf{(d)} Fitting each side of the peak in \textbf{(c)} to a form $A|N(\lambda - \lambda_c)|^{-\alpha} + C$, we plot the exponents $\alpha^\pm$ on the left and right as a function of $N$.
    }
\end{figure*}

Examining the von Neumann entropy $S_{vN}(t)$ and half-chain mutual info $I = 2S_{N/2} - S_A$ as a function of the ancilla coupling $\lambda$ in the star-Ising model ($J=0$), we again consider a quench from the effective infinite temperature initial product state $\ket{+y}^{\otimes N+1}$. We have numerically verified through the level statistics that the parameters we study lead to nonintegrable, random matrix statistics, so we expect the $y$-polarized state to obey ETH \cite{turner2018,pirmoradian_investigation_2025,pirmoradian_topological_2026,pal_many-body_2010} (For more on the level statistics, refer to the Supplemental Material).

With increasing coupling strength $\lambda$, we observe in Fig.\ref{fig:star-ising-scaling}(a) a surprising transition to a dynamics where the early-time super-ballistic dynamics for $\lambda t<2$ becomes sub-ballistic for times $\lambda t>2$; furthermore, in the strongly coupled limit $\lambda \gg h$ this scrambling time scale increases exponentially with $\lambda$. We compute the cumulative average half-chain mutual information
\begin{equation}
\bar{I}(t) \equiv \frac1{t} \int_0^t ds I(s)
\end{equation}
to eliminate some of the oscillatory behavior and fit to the average exponential form \eqref{eq:average_exp_fit} to study the scaling of the prethermal saturation tmie $\tau$ with the system parameters $\lambda, h, N$. We identify a crossover between the ancilla-enhanced entropy growth and ancilla-restricted behavior at a critical $\lambda_c$ through a divergent peak in the saturation time $\tau$ vs. $\lambda$ and study how $\lambda_c$ and the associated time scales evolve with $N$. An entropy growth rate which increases with system size indicates fast-scrambling behavior \cite{Belyansky2020}.

We observe three distinct plateaus in the average mutual information, shown in Fig.\ref{fig:star-ising-scaling}(b). For all parameters, the ancilla entropy $S_A$ saturates to $1$, while the half-chain entropy $S_{N/2}$ saturates to a value determined by the available Hilbert space. In general, the system obeys a permutation symmetry which restricts the spin chain to the Dicke manifold of states, whose size scales linearly with $N$\cite{latorre_entanglement_2005}. In the critical regime where $\lambda \sim h$ after the quench from $\ket{+y}^{\otimes N+1}$, the half-chain realizes a maximal entropy of $\frac1{2}(\log_2(N)+1) = 
\log_2(\sqrt{2N})$, giving a mutual information of $\log_2(N)$. When $\lambda \ll h$, the effective Hamiltonian derived before shows the system explores an ensemble of degenerate two-level systems, rather than a the full symmetric Hilbert space, which leads to the reduction in the mutual information by one bit. Finally, in the strongly coupled limit $\lambda \gg h$, the spin-chain forms an effective singlet with the ancilla, restricting the half-chain entropy and mutual information to $1$; the system only fully thermalizes to the mutual information bound of $\log_2(N)$ after a long time scale $\propto \lambda/h$. 

The saturation times to reach these plateaus, plotted vs. $\lambda N$ in Fig.\ref{fig:star-ising-scaling}(c) for different $N$, show a sharp divergence around the crossover $\lambda N/h\sim 8$. Normalizing the saturation times by $\sqrt{N}$ to reflect the scaling of the Hilbert space, we find the power law exponent of this divergence to be around $1$. The scaling of the saturation time with $\lambda, h$ for fixed $N$ is shown in the Supplemental info, with a scaling collapse of $\tau(\lambda)$ plotted vs. $\lambda/h$ and power law exponents $\approx 1$.

\section{\label{sec:RingAncilla} Ising Ring-Ancilla Model}

We now consider the full Ising Ring-Ancilla Hamiltonian Eq.~\eqref{eq:full_ring_star_ham}, initially setting $g=g_A=0$; the longitudinal field has a minimal effect in this model, since the ancilla coupling alone breaks the integrability of the Ising chain. The effective Hamiltonians formed by projecting onto the ancilla now include the full local dynamics of the interacting spin chain. In the basis formed by $\mathsf{P_x}$, the Hamiltonian takes the form of Eq.~\eqref{eq:bare_tfim_hamiltonian}, shifted by the ancilla field $\pm h_A/2$. Given the known eigenstates $\ket{n}$ of the TFIM $\mathsf{H_{TFIM}}\ket{n} = \varepsilon_n\ket{n}$, an initial state $\ket{n}\otimes \ket{+x}_A$ with energy $\varepsilon_n-h_A/2$ transforms under the perturbation into the state $\sum_m c_m \ket{m}\otimes\ket{-x}_A$ with $c_m = \bra{m}S^z\ket{n}$; therefore, a Feshbach resonance forms when $\omega_{nm} \equiv \varepsilon_n-\varepsilon_m = h_A$, where the energy absorbed or emitted by the ancilla spin flip exactly matches an energy gap of TFIM modes. The rate of these resonant transitions is given by Fermi's Golden Rule $\propto \lambda^2 \frac{|\bra{n}S^z\ket{m}|^2}{\omega_{nm}-h_A}$. Entanglement within the spin chain grows ballistically through dynamics of domain wall or magnon propagation, with the maximum speed of propagation with the spin chain at its critical point. Entropy growth between the ancilla and spin chain occurs at a rate $\sim \lambda^2$ due to dephasing between components in the $\ket{\pm x}_A$ sectors.

In the basis formed by $\mathsf{P_z}$, we instead have an effective Hamiltonian resembling a bare TFIM Eq.~\eqref{eq:bare_tfim_hamiltonian} with a longitudinal field term $\pm \lambda/2$ dependent on the ancilla state. This contributes a confining potential to the TFIM excitations, which slows entanglement growth within the spin chain in this phase. Including the longitudinal fields $g,g_A$ in the full Hamiltonian effectively renormalizes the confining potential experienced by the TFIM spin chain and modifies the energy gap associated with flipping the ancilla spin. In the extreme strong coupling limit $\lambda\gg J,h$, the system has an approximate permutation symmetry which restricts the Hilbert space to the symmetric Dicke subspace, leading to rapid, subextensive entropy growth of the collective spin chain degrees of freedom; the spin chain local degrees of freedom then slowly thermalize to realize the full, extensize entropy on a time scale $\propto 1/J$ dependent on the local permutation-symmetry-breaking term. The confining potential of the ancilla restricts early-time dynamics to sectors with fixed $S^z$, which imposes kinetic constraints on the motion of excitations and restricts the maximum entanglement growth.

\subsection{Mutual Information Dynamics}

\begin{figure*}
    \includegraphics[width=0.85\linewidth]{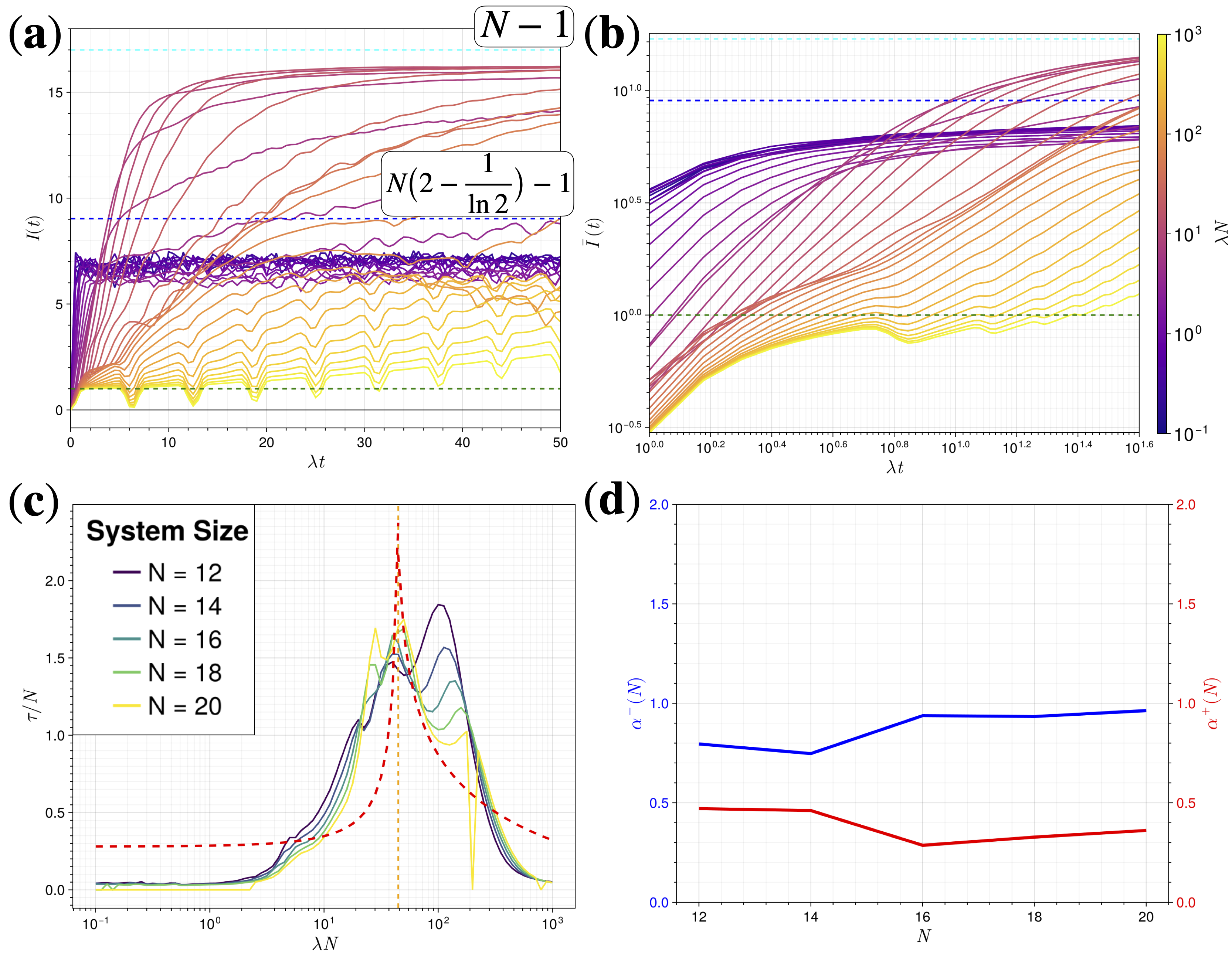}
    \caption{\label{fig:ising-ring-ancilla-MI-scaling}
    Entanglement growth results for the Ring-Ancilla model with $g=g_A=0$.
    \textbf{(a)} Half-chain mutual information for $N=20, \lambda \in[10^{-1},10^3], h=h_A=1.05$.
    \textbf{(b)} Average half-chain mutual information for $N=20, \lambda \in[10^{-1},10^3], h=h_A=1.05$; for $N\lambda \ll J,h$, we observe saturation to the same TFIM limit as before; when $\lambda \sim J,h$, we observe full spreading through the entire nonintegrable Hilbert space. Finally, for large $\lambda \gg J,h$, the ancilla confines the system to a singlet-like state with 1 bit of mutual information before slowly thermalizing to the Page limit. 
    \textbf{(c)} Fitting for the early time saturation and extracting the saturation time, we plot the saturation times $\tau/N$ against $\lambda N$ for even $N\in[12,20]$. We identify a critical slowing down and extract a power law exponent from either side of the peak (with example fitting curves shown dashed in red) with the exponents plotted vs $N$ in \textbf{(d)}. Here, the greatly enlarged Hilbert space of the fully nonintegrable model, compared to the Star Ising case, limited the range of times and system size available to our numerical techniques.
    }
\end{figure*}

Introducing the ancilla coupling $\lambda>0$ to a mixed field Ising chain maintains the nonintegrability and confining potential of the longitudinal field $g$; moreover, the shortest path between two chain sites is next-nearest neighbor always. This modifies the early-time entanglement growth as operators spread super-ballistically, characterized by the timescale $\sim 2/\lambda$ for spreading through this channel, where the factor of 2 counts the effective second-order process which mediates the next-nearest neighbor coupling. For very weak coupling $\lambda \ll J,h,g$, the ancilla Hilbert space is effectively negligible, acting as a cavity impurity.

Examining the mutual information quenches from $\ket{+y}^{\otimes N+1}$ in Fig.\ref{fig:ising-ring-ancilla-MI-scaling}(a,b), we can observe the effect that introducing the local coupling $J$ has on the available Hilbert space. For weak ancilla coupling $\lambda\ll J,h$, we recover the physics of the isolated TFIM from before, yielding a mutual information bound of $N(2-\frac1{\ln 2}-\frac1{N})$; however, the thermalization to this bound occurs much faster than for the isolated chain. For intermediate couplings, all terms compete in the Hamiltonian to realize a fully, nonintegrable limit of $N-1$ bits of mutual information. At large couplings $\lambda \gg J,h$, the local interaction $J$ commutes with the ancilla coupling, therefore leaving the entropy structure identical to that of the strongly coupled Star-Ising model before with $1$ bit of mutual information. The effective all-to-all coupling through the ancilla imposes a stricter constraint on the mobility of excitations in the chain than seen before in the low-field bare TFIM chain, altogether freezing the motion of domain walls rather than only preventing their creation/annhiliation. The extracted saturation times for the full ring-ancilla, normalized by system size, are plotted vs. $\lambda N$ in Fig. \ref{fig:ising-ring-ancilla-MI-scaling}(c), with the extracted power law exponents around the maximum plotted vs $N$ in Fig. \ref{fig:ising-ring-ancilla-MI-scaling}(d), suggesting an exponent $\approx 1$ on the left and $\approx 0.5$ on the right; however, the relatively enlarged Hilbert space of the nonintegrable, combined with the long saturation times near the crossover, preclude obtaining data of the same quality as the Star-Ising case.

The dynamic transition reveals that the ancilla can provide both a secondary channel for distributing entanglement and a severe restriction on entanglement growth depending on its coupling strength. The observed slow growth in the strongly coupled regime results completely from coherent effects, providing an example of disorder-free localization leading to diffusive, ``weak thermalization'' \cite{banuls_strong_2011}. We see a critical slowing down in the prethermal saturation time $\tau$ at the crossover point, while the time for saturation to the full nonintegrable Page limit continues to grow exponentially with $\lambda$; we thus observe rapid super-ballistic scrambling of collective DoF superimposed with slow, diffusive growth like $\sim\log(t)$ through the remaining Hilbert space. We explore this picture further by studying the operator spreading in the transverse and longitudinal OTOCs $(C_{xx},C_{zz})$.

\subsection{OTOC Dynamics}

\begin{figure*}
    \centering
    \includegraphics[width=0.85\linewidth]{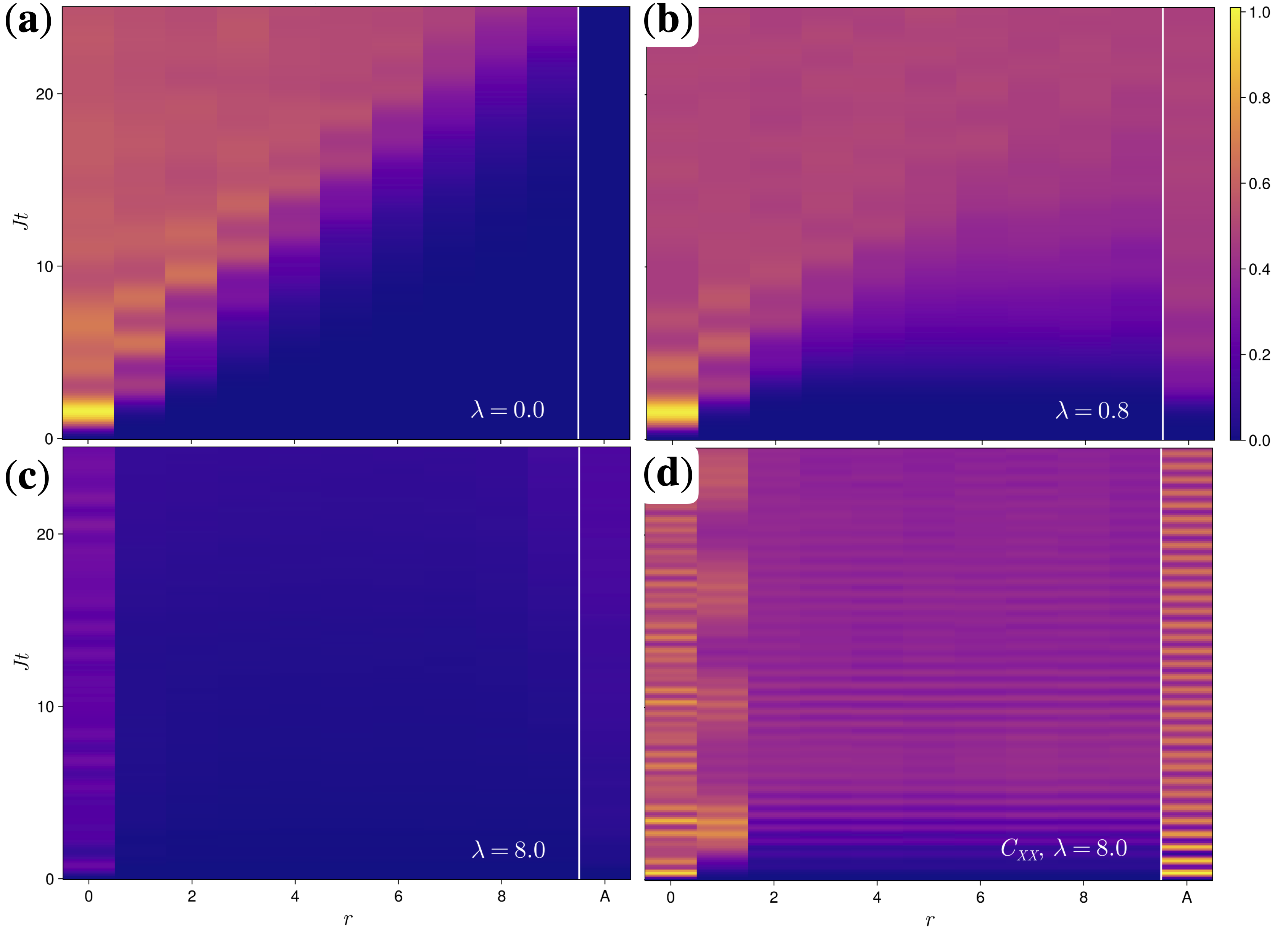}
    \caption{\label{fig:otoc_spreading}
    OTOC calculations for Ring-Ancilla model $J=1,h=h_A=1.05, g=g_A=0.45$ with open boundary (no bond $S_N^z S_1^z$)
    \textbf{(a)} $C_{zz}(r,t)$ for $\lambda=0$, showing the light cone spreading characteristic of the TFIM fermion excitations.
    \textbf{(b)} $C_{zz}(r,t)$ for $\lambda=0.8$, showing the uniform operator growth mediated through the ancilla across the entire chain, which provides a background to the light-cone spreading.
    \textbf{(c)} $C_{zz}(r,t)$ for $\lambda=8$, showing the operator confinement to the initial site with very slow growth on the ancilla.
    \textbf{(d)} $C_{xx}(r,t)$ for $\lambda=8$, showing rapid oscillations of operators transverse to the ancilla interaction on the ancilla and chain sites.
    }
\end{figure*}

Relaxing the restriction to begin in a product state, we approximate an infinite temperature state with a state selected randomly from the Haar measure and study the spread of operators under the influence of the Ring-Star Hamiltonian by explicitly calculating the OTOCs of the form Eq.~\eqref{eq:simplified_otoc_def} with ED, shown in Fig.~\ref{fig:otoc_spreading}. With $\lambda=0$, the local TFIM interactions lead to operator weight spreading along a ballistic light cone profile with a wavefront $\propto \frac{t^{2r}}{(r!)^2}$. With perturbatively small $\lambda$, the light cone structure remains apparent, with additional operator weight spreading uniformly in $r$ distributed through the ancilla super-ballistically on time scales $\sim 1/\lambda$. The growth timescale of the ancilla is half that of the sites in the chain, since $\sigma_z$ on the chain must decohere on the initial site and again on the central site to no longer commute with the local Ising interaction. In the strong coupling regime, a weak light cone persists but transfers smaller fractions of the original operator weight; the highly nonlocal ancilla coupling with increasing strength unintuitively restricts local operator spreading and prevents the light cone from spreading significantly. 

We gain analytical insight to this behavior from the Star-Ising autocorrelation Fig.~\ref{fig:autocorr}. The $N$-body interaction with the ancilla coherently projects operators onto $\sigma^z$ while aliasing orthogonal operators. As the initial $\sigma_i^z$ decoheres under the transverse field, it fails to commute with the Ising interaction, leading to operator weight growing on the ancilla and neighboring sites. When $\sigma^z$ grows support on the ancilla, it develops a coherent lifetime $\sim e^{-1/\lambda}$. Since $C_{zz}$ measures the failure of operators to commute with $\sigma^z$, we expect operators on the ancilla to strongly project onto the $\sigma^z$ space and coherently oscillate with frequency $\lambda/2$. As operator weight leaks from the ancilla to other chain sites, the same behavior from the ancilla (coherent oscillations and slow decay) imparts onto them. As the ancilla becomes highly entangled with the chain, this projection restricts how rapidly many-body operators on the spin chain decohere from $\sigma^z$, leading to slow behavior in the OTOC. Examining instead $C_{xx}$ in the strong coupling regime, shown in Fig.~\ref{fig:otoc_spreading}(d), we see strong oscillatory behavior, with a slow saturation to $\mathcal{O}(1)$ on all sites; the ancilla oscillations are stronger and $\pi$-phase shifted from the oscillations throughout the chain.

\begin{figure*}
    \includegraphics[width=0.85\linewidth]{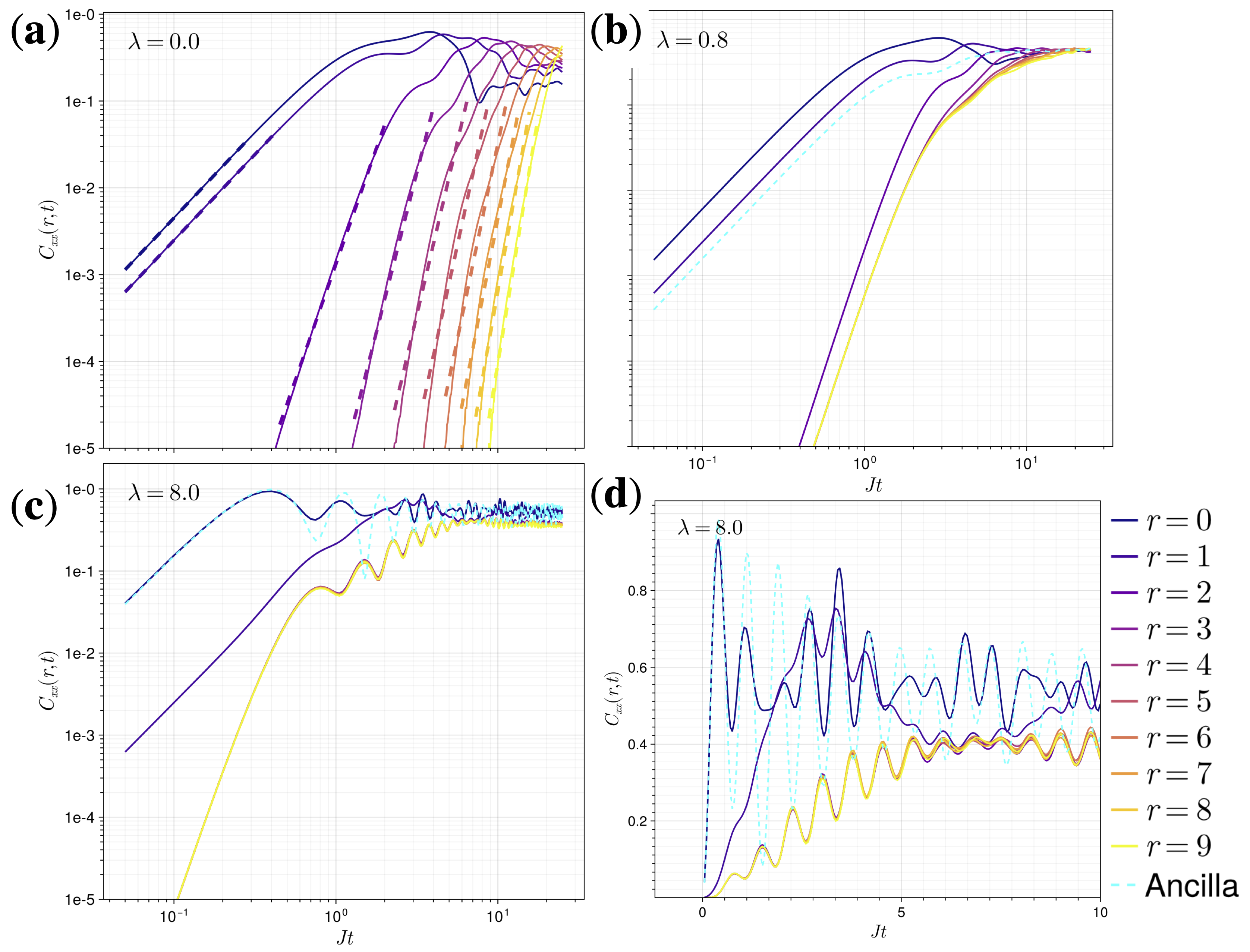}
    \caption{\label{fig:OTOC_traces_vs_r}
    Traces of $C_{xx}(r,t)$ for Ring-Ancilla model $J=1,h=h_A=1.05, g=g_A=0.45$ with open boundary (no bond $S_N^z S_1^z$)
    \textbf{(a)} At $\lambda=0$, we see evidence of the light-cone spreading for $r>1$, with polynomial growth in the OTOC up to its saturation; the slope of the polynomial growth weakly depends on $r$, while the saturation time increases with $r$.
    \textbf{(b)} At $\lambda=0.8$, the slopes at early times and saturation times for $r>2$  become nearly uniform, as the operator growth becomes dominated by the ancilla-mediated channel.
    \textbf{(c)} At $\lambda=8$, the OTOC uniformly saturates and oscillates as the ancilla coupling strongly pins $\sigma^z$ operators and rapidly drives orthogonal operators between the chain and ancilla.
    \textbf{(d)} $C_{xx}(r,t)$ at $\lambda=8$ with linear time-scale, carefully showing the rapid growth on the ancilla and ensuing oscillations on the bulk chain sites.
    }
\end{figure*}

\begin{figure*}
    \centering
    \includegraphics[width=0.85\linewidth]{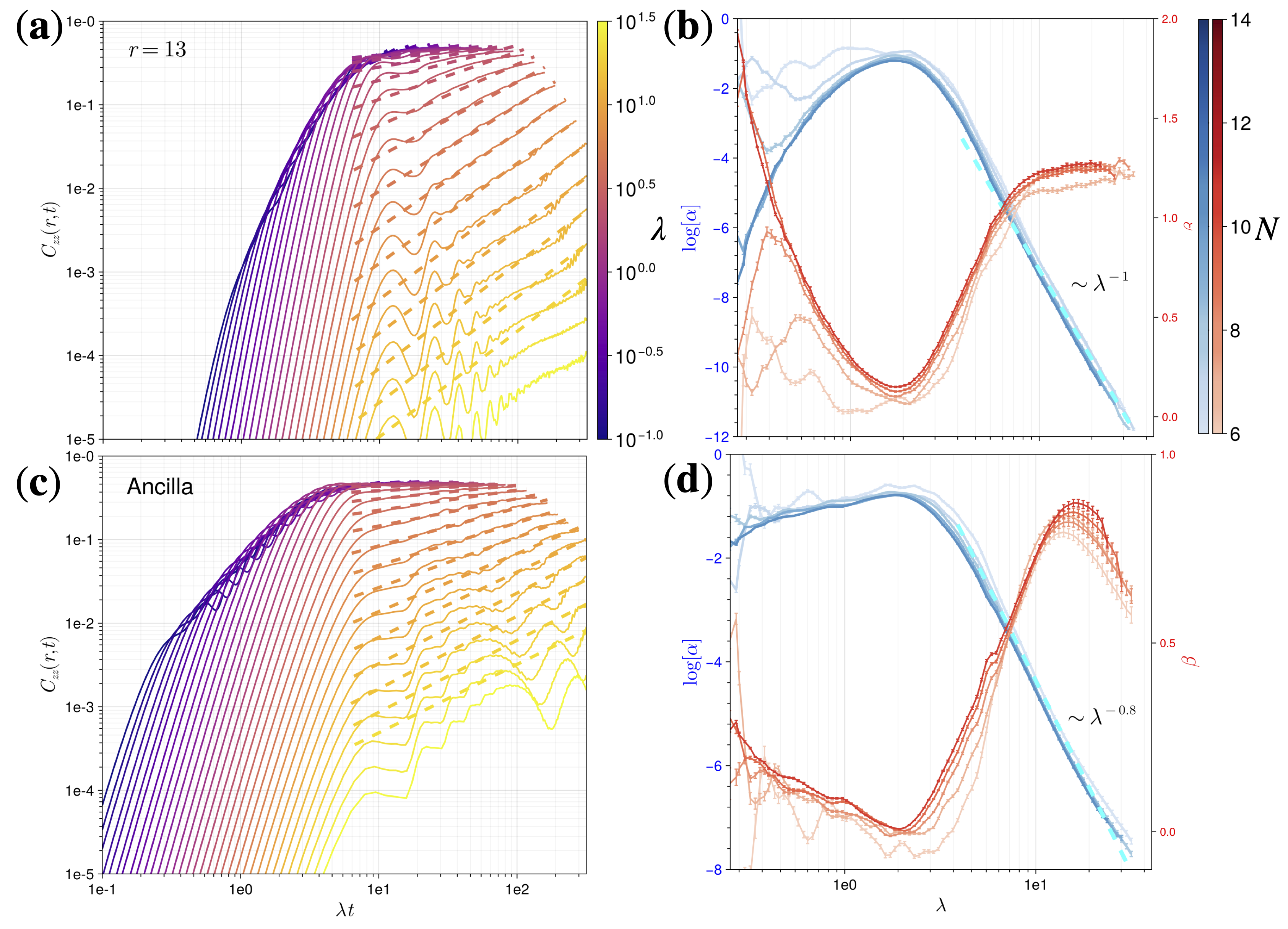}
    \caption{\label{fig:OTOC_traces_vs_lambda}
    \textbf{(a)} Traces of $C_{zz}(r=13 ; t)$ for $N=14,\lambda \in [0.1,10^{1.5}]$; the late time signal for each trace is fit to a form $\alpha t^\beta$.
    \textbf{(b)} Fit parameters for the traces of $C_{zz}(r=N-1,t)$ for $N\in[6,14]$; we find the strong coupling limit the amplitude $\log\alpha$ scales like $1/\lambda$.
    \textbf{(c)} Traces of $C_{zz}(r=A ; t)$ for $N=14,\lambda \in [0.1,10^{1.5}]$.
    \textbf{(d)} Fit parameters for the traces of $C_{zz}(r=A,t)$ for $N\in[6,14]$; we find the amplitude here scales like $\lambda^{-0.8}$ for large couplings.
    }
\end{figure*}

Individual traces at particular $r$ are plotted in Fig.~\ref{fig:OTOC_traces_vs_r}. The weak-coupling regime shows the polynomial growth associated with light-cone spreading on top of the slow, exponential growth mediated by the ancilla, while increasing $\lambda$ gradually eliminates the difference between traces $r>2$. The strong-coupling limit shown in Fig.~\ref{fig:OTOC_traces_vs_r}(d) demonstrates how $C_{xx}(A,t)$ quickly becomes $\mathcal{O}(1)$ in time $\pi/\lambda$, after which point the rate of change of $C_{xx}(r,t)$ within the chain descreases substantially. Once $\sigma_z$ exists on the ancilla, orthogonal operators rapidly fluctuate and experience an effective delay in their operator growth under the Ising interaction. Fig.~\ref{fig:OTOC_traces_vs_lambda} shows complementary perspectives from $C_{zz}(r=N-1,t)$ and $C_{zz}(A,t)$ for $N$ up to $14$. At early times $t<2\pi/\lambda$, the OTOC for finite $r$ scales $\propto h_A^2\lambda^4t^6$, following from the earlier BCH expansion. For times $t>2\pi/\lambda$, we fit the OTOC to the power law $\alpha t^\beta$ and find the coefficient $\alpha$ becomes exponentially suppressed in the strong-coupling limit, with $\log \alpha \propto \lambda^{-1}$, since the rate at which operator weight decoheres from the $\sigma_z$ space is approximately $e^{-\lambda}$. This slow growth for two-body operators outside the $\sigma_z$ space continues to be exponentially suppressed for larger many-body operators. The sublinear growth of the OTOCs into the bulk of the operator Hilbert space then generically provides sub-ballistic, diffusive entropy growth like $\sim \log t$, as observed in our quench simulations.

\section{\label{sec:Discussion} Discussion and Conclusions}

\begin{figure*}
    \includegraphics[width=0.85\linewidth]{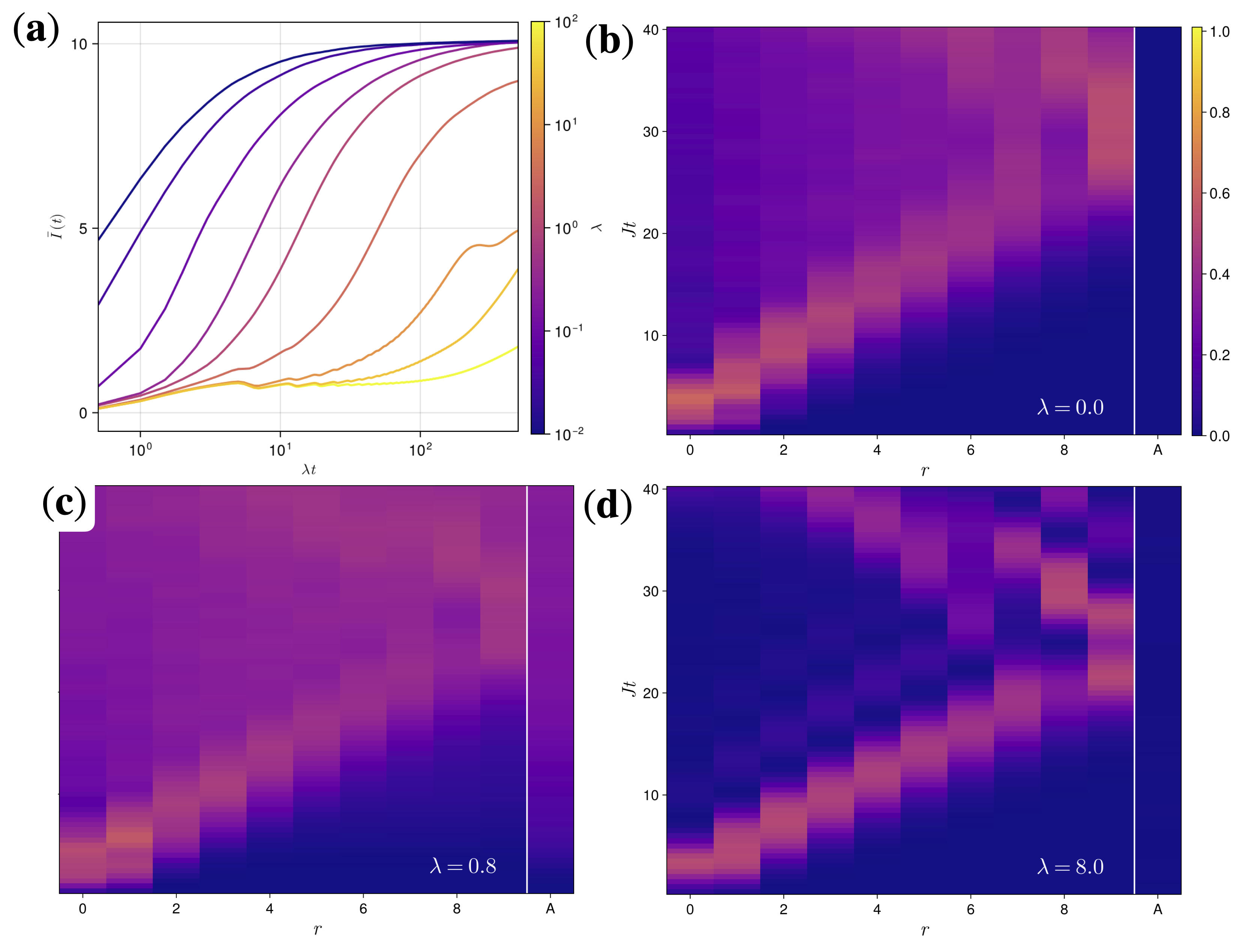}
    \caption{\label{fig:transverse_otocs}
    Results here shown for open-boundary ring-ancilla model with $N=14, J=1, h=1.05, g=0.45$ and transverse coupling term $\lambda S^x S_A^x$.
    \textbf{(a)} Average mutual information growth in time vs. $\lambda$; we see the same fast-to-slow transition as before for increasing coupling $\lambda$. 
    \textbf{(b)} $C_{xx}(r,t)$ for $\lambda =0$, showing expected ballistic propagation of operators.
    \textbf{(c)} $C_{xx}(r,t)$ for $\lambda=0.8$, showing ancilla-mediated all-to-all operator spreading.
    \textbf{(d)} $C_{xx}(r,t)$ for $\lambda=8$, demonstrating how the local Ising interaction $J S_j^z S_{j+1}^z$ allows for $x$-operator spreading, but the ancilla coupling $S^x S_A^x$ precludes it. As a result, the operator growth is confined to the edges of the light cone, rather than spreading uniformly across it.
    }
\end{figure*}
 
We have shown how tuning the coupling between a spin chain and an auxiliary qubit can interpolate between two opposite dynamical regimes: super-ballistic, ancilla-accelerated scrambling at weak coupling, and confined, sub-ballistic operator growth at strong coupling. This continuous crossover exists in a closed, time-independent Hamiltonian, controlled by a single dimensionless combination of microscopic parameters. We locate this crossover independently from two distinct diagnostics: a divergent peak in the saturation time of the half-chain mutual information at $\lambda_c N/h \simeq 8$ in the star-Ising limit, and an exponential suppression of the late-time OTOC growth rate at large coupling, $\log\alpha \propto 1/\lambda$. Independently identifying the same crossover in both an entanglement measure and an operator-space correlator, confirm this as a genuine dynamical transition rather than an artifact of either diagnostic.
 
Ref.~\cite{lucas2019quantum} found a that central ancilla coupling confines operators with coherent lifetime $\tau \propto h^2/(\lambda N)$ on a time-independent star-graph Hamiltonian; we sharpen that result in three distinct ways. First, restoring the local Ising coupling $J$ (Sec.~\ref{sec:RingAncilla}) puts the ancilla-mediated channel in direct competition with the chain's ballistic light cone rather than replacing it: the two channels coexist, and their competition sets the crossover from the bare TFIM entanglement bound to the fully nonintegrable $N-1$-bit volume law and back to a confined singlet-like state as $\lambda$ grows. Second, we explicitly identify the Landau-Zener physics originating the confinement: projecting out the ancilla shows that the effective energy gap for flipping its spin creates an avoided crossing each time the chain's magnetization sweeps through the resonance condition $\omega_{nm} = h_A$, and the resulting coherent destruction of tunneling between $S^z$ sectors drives the observed prethermal oscillations in the entropy. This provides a dynamical, rather than purely phenomenological, account of why the entanglement growth stalls. Finally, we show confinement does not depend on the specific $S^zS^z$ form of the coupling: by switching the ancilla to couple through $S^xS^x$ instead, as demonstrated in Fig.~\ref{fig:transverse_otocs}, operators become confined to the edge of the light cone rather than freezing them outright, as $\sigma^x$ operator strings can spread through the local channel, but not through the ancilla. This illustrates the confinement effect as a generic feature of any global coupling axis, not solely the fine-tuned $S^zS^z$ interaction initially considered.
 
Altogether, we have identified a static, disorder-free route to slow, non-ergodic dynamics without the conventional ingredients of disordered many-body localization or periodic Floquet driving. Our model also differs from the fast-scrambling behavior reported for RUC realizations of the same star geometry \cite{lucas2019quantum, Harrow2021}, since here the identical geometry realized as a time-independent Hamiltonian produces the opposite, confined regime once the coupling exceeds $\lambda_c$. The slow dynamics instead result from a single coherent, global coupling whose own entanglement with the chain saturates quickly enough to project the chain's operators into effectively decoupled subspaces; this mechanism avails itself to any system with a bottlenecked auxiliary degree of freedom, without requiring disorder, driving, or an extensively engineered interaction graph. By tuning the ancilla coupling, one controls whether operators become frozen in the highly-entangled ancilla or globally distributed throughout the spin system. 
 
The bare spin chain Hamiltonian considered here is integrable at $g=0$ but generically nonintegrable otherwise; the confinement transition itself appears robust to this distinction, since the ancilla coupling naturally breaks the integrability of the $g=0$ model, and the crossover survives with $g\neq0$. This raises the questions of whether the same transition can be tuned by integrability directly (for example in spin-conserving chains coupled to an ancilla, or by explicit disorder in the bonds or local fields) and whether an analogous transition appears in RUC realizations by treating the ancilla coupling as a tunable strength rather than fixing it at the fast-scrambling limit \cite{kuriyattil2023onset}.
 
Beyond this, two natural extensions follow from treating the ancilla as a genuine open-system handle. In contrast to integrating out the ancilla to obtain effective non-Hermitian dynamics on the chain, one could ask whether particular non-Hermitian problems admit a unitary embedding with a hidden qubit of exactly this kind \cite{singh2024embedding}. One could also examine how the transition scales with the ancilla's Hilbert space dimension, since enlarging the ancilla spin toward the classical limit approaches the analog of a bosonic cavity mode; this invites the possibility of coupling a qubit array to a cavity or other auxiliary degree of freedom to deliberately protect quantum simulators against environmental dephasing \cite{Ng_mbl2019,diniz2011strongly,putz2014protecting}.

\appendix*
\section{Data Availability}
Data and codes are available through Zenodo \cite{zenodo_repo}.

\begin{acknowledgments}
This material is based upon work supported by the National Science Foundation (NSF) NRT-QISE DGE-2244045 (RB) and NSF DMR-2138905 (JS, NT). Computations were done using a combination of python QuSpin \cite{quspin} and Julia KrylovKit \cite{Haegeman_KrylovKit_2024} packages on the Unity Cluster at The Ohio State University. 
\end{acknowledgments}

\bibliography{refs}

\end{document}